\documentclass{article}

\usepackage{PRIMEarxiv}
\usepackage[frozencache,cachedir=.]{minted}
\usepackage{xpatch}
\xpatchcmd{\mintinline}{\begingroup}{\begingroup}{}{}
\xpatchcmd{\minted}{\VerbatimEnvironment}{\VerbatimEnvironment}{}{}
\usepackage[utf8]{inputenc} 
\usepackage[T1]{fontenc}    
\usepackage{hyperref}       
\usepackage{url}            
\usepackage{booktabs}       
\usepackage{amsfonts}       
\usepackage{nicefrac}       
\usepackage{microtype}      
\usepackage{lipsum}
\usepackage{fancyhdr}       
\usepackage{graphicx, adjustbox}
\usepackage{verbatim,listings}
\usepackage{xcolor,thumbpdf,lmodern,rotating}
\usepackage{amsfonts,amssymb,pifont}
\usepackage{amsmath,mathtools}
\usepackage{caption}
\usepackage{subcaption}
\usepackage{url}
\usepackage{cleveref}
\usepackage{float}
\usepackage{multirow}
\graphicspath{{images/}} 
\usepackage[]{apacite}
\usepackage{makecell}
\usepackage{algorithm}
\usepackage{algpseudocode}
\usepackage{mathrsfs}

\usepackage{siunitx}

\usepackage{caption} 
\usepackage{tikz}
\usetikzlibrary{decorations.pathreplacing}

\definecolor{royalblue}{rgb}{0.25, 0.41, 0.88}
\definecolor{darkcandyapplered}{rgb}{0.64, 0.0, 0.0}

\title{GEMAct: a Python package for non-life (re)insurance modelling
}

\author{
  Gabriele Pittarello \\
  Universit\`{a} `La Sapienza'\\
  Roma\\
  \texttt{gabriele.pittarello@uniroma1.it} \\
   \And
 Edoardo Luini\\
 Universit\`{a} Cattolica del Sacro Cuore\\
  Milano\\
  \texttt{edoardo.glaucoluini@unicatt.it} \\
     \And
  Manfred Marvin Marchione\\
 Universit\`{a} `La Sapienza'\\
  Roma\\
  \texttt{manfredmarvin.marchione@uniroma1.it} \\
}

\begin{document}
\maketitle

\begin{abstract}
This paper introduces \textbf{gemact}, a \textbf{Python} package for actuarial modelling based on the collective risk model. The library supports applications to risk costing and risk transfer, loss aggregation, and loss reserving. We add new probability distributions to those available in \textbf{scipy}, including the (a, b, 0) and (a, b, 1) discrete distributions, copulas of the Archimedean family, the Gaussian, the Student t and  the Fundamental copulas.
We provide an implementation of the AEP algorithm for calculating the cumulative distribution function of the sum of dependent, non-negative random variables, given their dependency structure specified with a copula. The theoretical framework is introduced at the beginning of each section to give the reader with a sufficient understanding of the underlying actuarial models.
\end{abstract}

\keywords{insurance, collective risk model, risk costing, loss aggregation, claims reserving, Python}

\section[Introduction]{Introduction} \label{sec:intro}

In non-life insurance, the accurate representation and quantification of future losses is a foundational task, central to several areas ranging from pricing and reserving to risk management. Indeed, the actuarial literature is rich in models that are relevant in such applications. Among those, the collective risk model has been widely studied as it is mathematically tractable, it requires little and general information, and it can be efficiently implemented \cite{klugman98, embrechts09, parodi14}. In particular, by knowing the frequency and severity distributions of the losses, it is possible to compute the distribution of the aggregate (total) losses. 
The collective risk model represents the common thread of this work and we developed \textbf{gemact} to provide a collection of tools for (re)insurance modelling under a unified formal framework.

After a brief discussion on how to install the software in \Cref{sec:installation}, we introduce the statistical framework of the collective risk model in \Cref{sec:lossmodel}. 
There, we define an aggregate loss distribution as a random sum of i.i.d. random variables, which can be computed using the recursive formula \cite{panjer81}, the discrete Fourier transform \cite{buhlmann84, wang98, grubel99} and a Monte Carlo simulation approach \cite[p.~467]{klugman98}. Once the aggregate loss distribution is available, its expected value can be used for costing purposes. In this respect, the package supports typical coverage modifiers like (individual and aggregate) deductibles, limits, and reinstatements \cite<see>{Sundt90}. Also, we consider different methods for the discretisation of continuous distributions \cite{gerber82}. 

Often, it is necessary to model the sum of a fixed number of dependent random variables. In order to do so, in \Cref{sec:lossaggregation}, we introduce the AEP algorithm \cite{arbenz11} and a Monte Carlo simulation approach for evaluating the cumulative distribution function of a sum of random variables with a given dependency structure. The dependency structure can be specified with the copulas we implemented. These are listed in \Cref{app:alldists} and include copulas of the Archimedean family, the Gaussian, the Student t and the Fundamental copulas \cite{nelsen07}.

Lastly, assuming a collective risk model holds for the cells of a loss development triangle, it is possible to define the stochastic claims reserving model in \citeA{ricotta16, clemente19}. In this case, the user obtains information on the frequency and severity parameters of the cells from the Fisher-Lange method \cite{fisher99}. Both these approaches are described in \Cref{sec:lossreserve}.

\subsection{Context, scope, and contributions} 

In the recent years, programming languages and statistical computing environments, such as \textbf{Python} \cite{python} and \textbf{R} \cite{R}, have become increasingly popular \cite{ozgur22}. Currently, coding skills form part of the body of knowledge of actuaries and actuarial science researchers.
In \textbf{R}, an extensive implementation for aggregate loss modelling based on the collective risk theory is the \textbf{actuar} package \cite{actuar1, actuar2}.
An available library in Python is \textbf{aggregate}, which implements the computation of compound probability distributions via fast Fourier transform convolution algorithm \cite{aggregatepackage}. This package employs a specific grammar for the user to define insurance policy declarations and distribution calculation features. Direct access to its objects and their components is also possible.

With regards to claims reserving, \textbf{chainladder} offers in \textbf{Python} standard aggregate reserving techniques, like deterministic and stochastic chain-ladder methods, the Bornhuetter-Ferguson model, and the Cape Cod model \cite{chainladderPypackage}. This package is available in \textbf{R} and \textbf{Python} \cite{chainladderR}. Furthermore, \textbf{apc} provides the family of age-period-cohort approaches for reserving. This package is also available in both the above-mentioned programming languages \cite{apcpackage}. 

When it comes to the topic of dependence modelling via copulas, in \textbf{Python} one can use the \textbf{copulas} and \textbf{copulae} packages \cite{copulaepackage, copulaspackage}. Similarly, copula features in \textbf{R} are implemented in \textbf{copula}, see the package and its extensions in \citeA{copulaR1, copulaR2, copulaR3}. 

In this manuscript, we present an open-source \textbf{Python} package that extends the existing software tools available to the actuarial community.
Our work is primarily aimed at the academic audience, who can benefit from our implementation for research and teaching purposes. Nonetheless, our package can also support non-life actuarial professionals in prototypes modelling, benchmarking and comparative analyses, and ad-hoc business studies. 

From the perspective of the package design, 
\textbf{gemact} adopts an explicit, direct and integrated object-oriented programming (OOP) paradigm. In summary, our goal is to provide: 
\begin{itemize}
    \item 
    A computational software environment that gives users control over mathematical aspects and actuarial features, enabling the creation of models tailored to specific needs and requirements. 
    \item 
    An object oriented system whose elements (i.e. objects, methods, attributes) 
    can be accessed and managed via our API in such a way as to be interactive and suitable for users familiar with OOP designs and with the underlying modelling framework.
    \item A collection of extensible libraries to make \textbf{gemact} survive over time. Our package is designed in an attempt to be easily extended or integrated with new functionalities, modules, and to respect the attributes that qualify extensible software \cite{johansson09}. Namely, a modification to the functionalities should involve the least number of changes to the least number of possible elements \cite<modifiability,>[p.~137]{bass03}, the addition of new requirements should not raise new errors \cite<maintainability,>[p.~25]{sommerville11}, and the system should be able to expand in a chosen dimension without major modifications to its architecture \cite<scalability,>{bondi00}.
\end{itemize}
From the perspective of actuarial advancements and developments, \textbf{gemact} provides an implementation to established algorithms and methodologies. In particular, our package:
\begin{itemize}
    \item Implements (a, b, 0) and (a, b, 1) classes of distributions for describing the loss frequency \cite[p.~505]{klugman98}, and further continuous distributions to model the loss severity, like the generalised beta \cite[p.~493]{klugman98}. Additional details can be found in \Cref{app:alldists}. Moreover, it integrates these into \textbf{scipy} distributions \cite{scipy}.
    \item Offers the first open-source software implementation of the AEP algorithm \cite{arbenz11} for evaluating the cumulative distribution function of a sum of random variables with a given dependency structure specified via a copula.
    \item Includes the Student t copula and a method for numerically approximating its cumulative distribution function \cite{GenzBretz1999, GenzBretz2002}.
    \item Implements the stochastic claims reserving model described by \citeA{ricotta16, clemente19} based upon the collective risk model apparatus.
\end{itemize}

\section{Installation}
\label{sec:installation} 

The production version of the package is available on the Python Package Index (\textbf{PyPi}).
Users can install \textbf{gemact} via \textbf{pip}, using the following command in the operating system command line interface.

\begin{minted}[mathescape]{console}
pip install gemact==1.2.1
\end{minted}

Examples on how to get started and utilise objects and functionalities of our package will be shown below. This work refers to production version \texttt{1.2.1}.

Furthermore, the developer version of \textbf{gemact} can be found on \textbf{GitHub} at:

\centerline{\href{https://github.com/gpitt71/gemact-code}{https://github.com/gpitt71/gemact-code}.}

Additional resources on our project, including installation guidelines, API reference, technical documentations and illustrative examples, can be found at:

\centerline{\href{https://gem-analytics.github.io/gemact/}{https://gem-analytics.github.io/gemact/}.}

\section{Loss model}
\label{sec:lossmodel}

Within the framework of the collective risk model \cite{embrechts09}, all random variables are defined on some fixed probability space $(\Omega, \mathcal{F}, P)$. Let
\begin{itemize}
    \item $N$ be a random variable taking values in $\mathbb{N}_0$ representing the claim frequency.
    \item $\left\{ Z_i\right\}_{i \in \mathbb{N}}$ be a sequence of i.i.d non-negative random variables independent of $N$; $Z$ is the random variable representing the individual (claim) loss.
\end{itemize}

The aggregate loss $X$, also referred to as aggregate claim cost, is

\begin{equation}
\label{eq:lossmodel}
X=\sum_{i=1}^{N} Z_i,
\end{equation}

with $\sum_{i=1}^{0} Z_i=0$. Details on the distribution functions of $N$, $Z$ and $X$ are discussed in the next sections.
\Cref{eq:lossmodel} is often referred to as the frequency-severity loss model representation. This can encompass common coverage modifiers present in (re)insurance contracts \cite[p.~50]{parodi14}. More specifically, let us consider:

\begin{itemize}

    \item For $a \in [0, 1]$, the function $Q_a$ apportioning  the aggregate loss amount: 
    \begin{equation}
        \label{eq:qs}
       Q_a (X)= a X.
    \end{equation}
    
    \item  For $c,d \geq 0$, the function $L_{c, d}$ applied to the individual claim loss: 
    
     \begin{equation}
         \label{eq:minmax}
             L_{c, d} (Z_i) = \min \left\{\max \left\{0, Z_i-d\right\}, c\right\}. 
         \end{equation}
    
    Herein, for each and every loss, the excess to a \textit{deductible} $d$ (sometimes referred to as \textit{priority}) is considered up to a \textit{cover} or \textit{limit} $c$. In line with \citeA[p.~34]{albrecher17}, we denote $[d, d+c ]$ as \textit{layer}. An analogous notation is found in \citeA{ladoucette06} and \citeA[p.~46]{parodi14}. Similarly to the individual loss $Z_i$, \Cref{eq:minmax} can be applied to the aggregate loss $X$.

\end{itemize}

Computing the aggregate loss distribution is relevant for several actuarial applications \cite[p.~93]{parodi14}. The \textbf{gemact} package provides various methods for calculating the distribution of the loss model in \Cref{eq:lossmodel} that allow the transformations of \Cref{eq:qs}, \Cref{eq:minmax} and their combinations to be included.

\subsection{Risk costing}

In this section, we describe an application of the collective risk model of \Cref{eq:lossmodel}. The expected value of the aggregate loss of a portfolio constitutes the building block of an insurance tariff. This expected amount is called pure premium or loss cost and its calculation is referred as \textit{risk costing} \cite[p.~282]{parodi14}. Insurers frequently cede parts of their losses to reinsurers, and risk costing takes this transfers into account. Listed below are some examples of basic reinsurance contracts whose pure premium can be computed using \textbf{gemact}.

\begin{itemize}
    \item The \textit{Quota Share (QS)}, where a share $a$ of the aggregate loss ceded to the reinsurance (along with the respective premium) and the remaining part is retained:
    
    \begin{equation}
        \text{P}^{QS} = \mathbb{E}\left[ Q_a \left( X \right)\right].
    \end{equation}
    
    \item The \textit{Excess-of-loss (XL)}, where the insurer cedes to the reinsurer each and every loss exceeding a deductible $d$, up to an agreed limit or cover $c$, with $c,d \geq 0$: 
    \begin{equation}
        \text{P}^{XL} =  \mathbb{E}\left[ \sum_{i=1}^{N} L_{c,d} (Z_i) \right].
    \end{equation}
    
\item The \textit{Stop Loss (SL)}, where the reinsurer covers the aggregate loss exceedance of a (aggregate) deductible $v$, up to a (aggregate) limit or cover $u$, with $u,v \geq 0$:
    \begin{equation}
    \label{eq:sl}
        \text{P}^{SL} = \mathbb{E}\left[ L_{u, v} (X) \right].
    \end{equation} 
\end{itemize}

The model introduced by \Cref{eq:lossmodel} and implemented in \textbf{gemact} can be used for costing contracts like the \textit{Excess-of-loss with Reinstatements (RS)} in \citeA{Sundt90}. Assuming the aggregate cover $u$ is equal to $ (K + 1) c $, with $K \in \mathbb{Z}^+$:

\begin{equation}
\label{eq:reinstatementsP}
    \text{P}^{RS} =  \frac{\mathbb{E}\left[ L_{u, v} (X) \right]}{1+\frac{1}{c} \sum_{k=1}^K l_k \mathbb{E}\left[ L_{c, (k-1)c+v}(X) \right]},
\end{equation}

where $K$ is the number of reinstatement layers and
$l_k \in [0, 1]$ is the reinstatement premium percentage, with $k=1, \ldots, K$.
When $l_k = 0$, the $k$-th resinstatement is said to be free.
In detail, the logic we implemented implies that whenever a layer is used the cedent pays a reinstatement premium, i.e. $l_k \text{P}^{RS}$, and the cover $c$ is thus reinstated. The reinstatement
premium will usually be paid in proportion to the amount that needs to be reinstated \cite[p.~52]{parodi14}.
In practice, the reinstatement premium percentage $l_k$ is a contractual element, given a priori as a percentage of the premium paid for the initial layer. In fact, in \textbf{gemact} $l_k$ is provided by the user.
The mathematics behind the derivation of $\text{P}^{RS}$ is beyond the scope of this manuscript; the interested reader can refer to \citeA{Sundt90}, \citeA[p.~325]{parodi14} and \citeA[p.~57]{Antal09}.

\subsection{Computational methods for the aggregate loss distribution}
\label{sec:computationalmethods}

The cumulative distribution function (cdf) of the aggregate loss in \Cref{eq:lossmodel} is:
\begin{align}
\label{eq:lossmodelcdf}
    F_X(x)=P[X \leq x]=\sum_{k=0}^{\infty} p_k F_Z^{* k}(x)
\end{align}
where $p_k=P[N=k]$, $F_Z(x)=P\left[ Z \leq x \right]$ and $F_Z^{* k}(x)=P\left[ Z_1 + \ldots + Z_n \leq x \right]$. 

Moreover, the characteristic function of the aggregate loss $\phi_{X}: \mathbb{R} \rightarrow \mathbb{C}$ can be expressed in the form:

\begin{equation}\label{eq:aggchar}\phi_{X}(t)=\mathcal{P}_{N}\left(\phi_{Z}(t)\right),\end{equation}

where $\mathcal{P}_{N}(t)=\mathrm{E}\left[t^N\right]$ is the probability generating function of the frequency $N$ and  $\phi_{Z}(t)$ is the characteristic function of the severity $Z$ \cite[p.~153]{klugman98}.

The distribution in \Cref{eq:lossmodelcdf}, except in a few cases, cannot be computed analytically and its direct calculation is numerically expensive \cite[p.~239]{parodi14}. For this reason, different approaches have been analysed to approximate the distribution function of the aggregate loss including parametric and numerical quasi-exact methods \cite<for a detailed treatment refer to>{Shevchenko:2010}. Amongst the latter, \textbf{gemact} implements \textit{Monte Carlo simulation} \cite[p.~467]{klugman98}, \textit{discrete Fourier transform} \cite{buhlmann84, wang98, grubel99} and the so-called \textit{recursive formula} \cite{panjer81}. A brief comparison of accuracy, flexibility and speed of these methods can be found in \citeA[p.~260]{parodi14} and \citeA[p.~127]{MVW2023}.
This section details these last two computational methods based on discrete mathematics to approximate the aggregate loss distribution. 

Henceforth, let us consider, for $j=0,1,2,\ldots,m-1$ and $h>0$, an arithmetic severity distribution with probability sequence:

\begin{align*}
    \mathbf{\{f\}}=\{f_0, f_1, \ldots , f_{m-1}\},
\end{align*}

where $f_j=P[Z=j\cdot h]$. 
The discrete version of \Cref{eq:lossmodelcdf} becomes

$$
g_s=\sum_{k=0}^{\infty} p_k f_s^{* k},
$$

where $g_s = P[X=s]$ and 

\[
f_s^{* j}:= \begin{cases}1 & \text { if } k=0 \text { and } s=0 \\ 0 & \text { if } k=0 \text { and } s \in \mathbb{N} \\ \sum_{i=0}^s f_{s-i}^{*(k-1)} f_i & \text { if } k > 0.\end{cases}
\]

\subsubsection{Discrete Fourier transform}
\label{sub:DFT}

The discrete Fourier transform (DFT) of the severity $\mathbf{\{f\}}$ is, for $k=0,...,m-1$, the sequence

$$\mathbf{\{\hat{f}\}}= \{\hat{f}_0, \hat{f}_1, \ldots, \hat{f}_{m-1} \},$$

where

\begin{equation}
\label{eq:DFT}
\widehat{f}_k=\sum_{j=0}^{m-1}f_j e^{\frac{2\pi i k j}{m}}.
\end{equation}

The original sequence can be reconstructed with the inverse DFT:

$$f_j=\frac{1}{m} \sum_{k=0}^{m-1} \widehat{f_k} e^{-\frac{i 2 \pi j k}{ m}}.$$

The sequence of probabilities $\mathbf{\{g\}}=\{g_0, g_1, \ldots, g_{m-1}\}$ can be approximated taking the inverse DFT of 

\begin{equation}
\mathbf{\{ \hat{g}\}}\coloneqq \mathcal{P}_{N}\left(\mathbf{\{\hat{f}\}}\right).
\end{equation}

The original sequence can be computed efficiently with a fast Fourier transform (FFT) algorithm, when $m$ is a power of $2$ \cite{embrechts09}.

\subsubsection{Recursive formula}
\label{ss:recursiveformula}

Assume that the frequency distribution belongs to the $(a, b, 0)$ class, i.e. for $k \geq 1$ and $a, b\in\mathbb{R}$:

\begin{equation}
    \label{eq:ab0def}
    p_k=\left(a+\frac{b}{k}\right)p_{k-1}.
\end{equation}

Here, $p_0$ is an additional parameter of the distribution \cite[p.~505]{klugman98}. The $(a, b, 0)$ class can be generalised to the $(a, b, 1)$ class assuming that the recursion in \Cref{eq:ab0def} holds for $k=2,3,\ldots$

The recursive formula was developed to compute the distribution of the aggregate loss when the frequency distribution belongs to the $(a,b,0)$ or the $(a,b,1)$ class. The sequence of probabilities $\mathbf{\{{g}\}}$ can obtained recursively using the following formula:

\begin{equation}
\label{eq:rec}
    g_s=\frac{\left[p_1-(a+b) p_0\right] f_s+\sum_{j=1}^s(a+b j / s) f_j g_{s-j}}{1-a f_0},
\end{equation}

with $1 \leq s \leq m-1$ and given the initial condition $g_0=\mathcal{P}_{N}\left(f_0\right)$.

\subsection{Severity discretisation}
\label{sec:severitydiscretisation}

The calculation of the aggregate loss with DFT or with the recursive formula requires an arithmetic severity distribution \cite{embrechts09}. Conversely, the severity distribution in \Cref{eq:lossmodel} is often calibrated on a continuous support. In general, one needs to choose a discretisation approach to arithmetise the original distribution. This section illustrates the methods for the discretisation of a continuous severity distribution available in the \textbf{gemact} package.

Let $F_z: \mathbb{R^+} \rightarrow [0, 1]$ be the cdf of the distribution to be discretised. 
For a bandwidth, or discretisation step, $h>0$ and an integer $m$, a probability $f_j$ is assigned to each point $h j$, with $j= 0, \ldots ,m-1$. Four alternative methods for determining the values for $f_j$ are implemented in \textbf{gemact}.

\begin{enumerate}
    \item The method of \textit{mass dispersal}:

    \begin{align*}
    f_j= \begin{cases}F_Z \left( \frac{h}{2} \right), & j=0 \\ F_Z\left( hj+ \frac{h}{2} \right)-F_Z\left(hj-\frac{h}{2}\right) &  j=1,\ldots,m-2\\
     1-F_Z\left(hj-\frac{h}{2}\right) &  j=m-1. \end{cases}
    \end{align*}

\item The method of the \textit{upper discretisation}:

    \begin{align*}
    f_j= \begin{cases}F_Z\left( hj+ h \right)-F_Z\left(hj\right) &  j=0,\ldots,m-2\\
     1-F_Z\left(hj\right) &  j=m-1. \end{cases}
    \end{align*}

    \item The method of the \textit{lower discretisation}:

    \begin{align*}
    f_j= \begin{cases}F_Z \left( 0 \right), & j=0 \\ F_Z\left( hj\right)-F_Z\left(hj-h\right) &  j=1,\ldots,m-1.\\
    \end{cases}
    \end{align*}

    \item The method of \textit{local moment matching}: 
    
    \begin{align*}
    f_j= \begin{cases}1-\frac{\mathrm{E}[Z\wedge h] }{h}, & j=0 \\ \frac{2 \mathrm{E}[Z \wedge h j]-\mathrm{E}[Z \wedge h (j-1) ]-\mathrm{E}[Z \wedge h (j+1) ]}{h} &  j=1,\ldots,m-1\end{cases}
    \end{align*}

    where $\mathrm{E}[Z\wedge h] = \int_{-\infty}^h t \; dF_z(t)+h[1-F_z(h)]$.
    
\end{enumerate}

\begin{figure}
\centering
\begin{subfigure}[b]{0.45\textwidth}
    \centering\includegraphics[width=8cm]{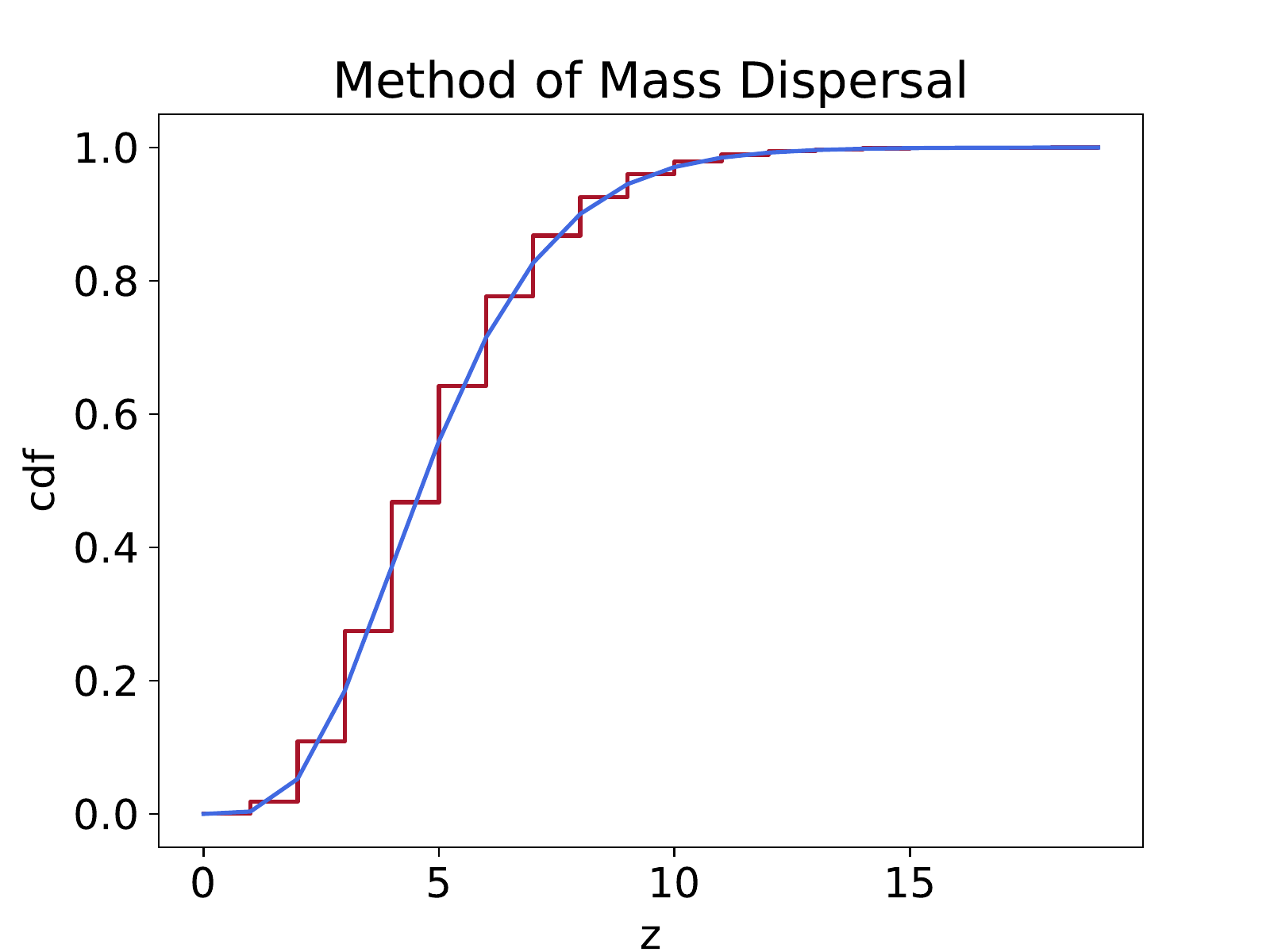}
\end{subfigure}
\hfill
\begin{subfigure}[b]{0.45\textwidth}
    \centering\includegraphics[width=8cm]{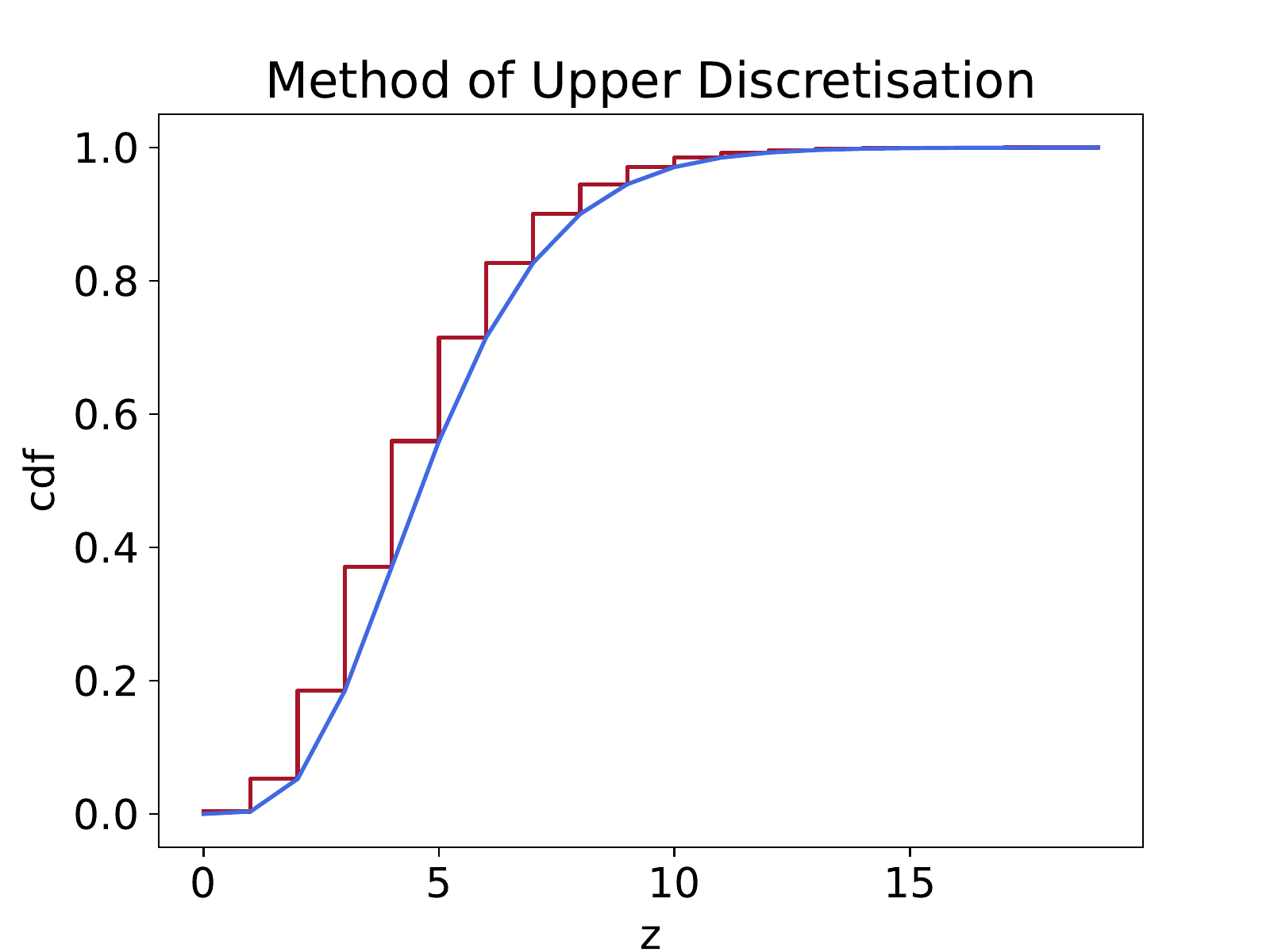}
\end{subfigure}
\hfill
\begin{subfigure}[b]{0.45\textwidth}
    \centering\includegraphics[width=8cm]{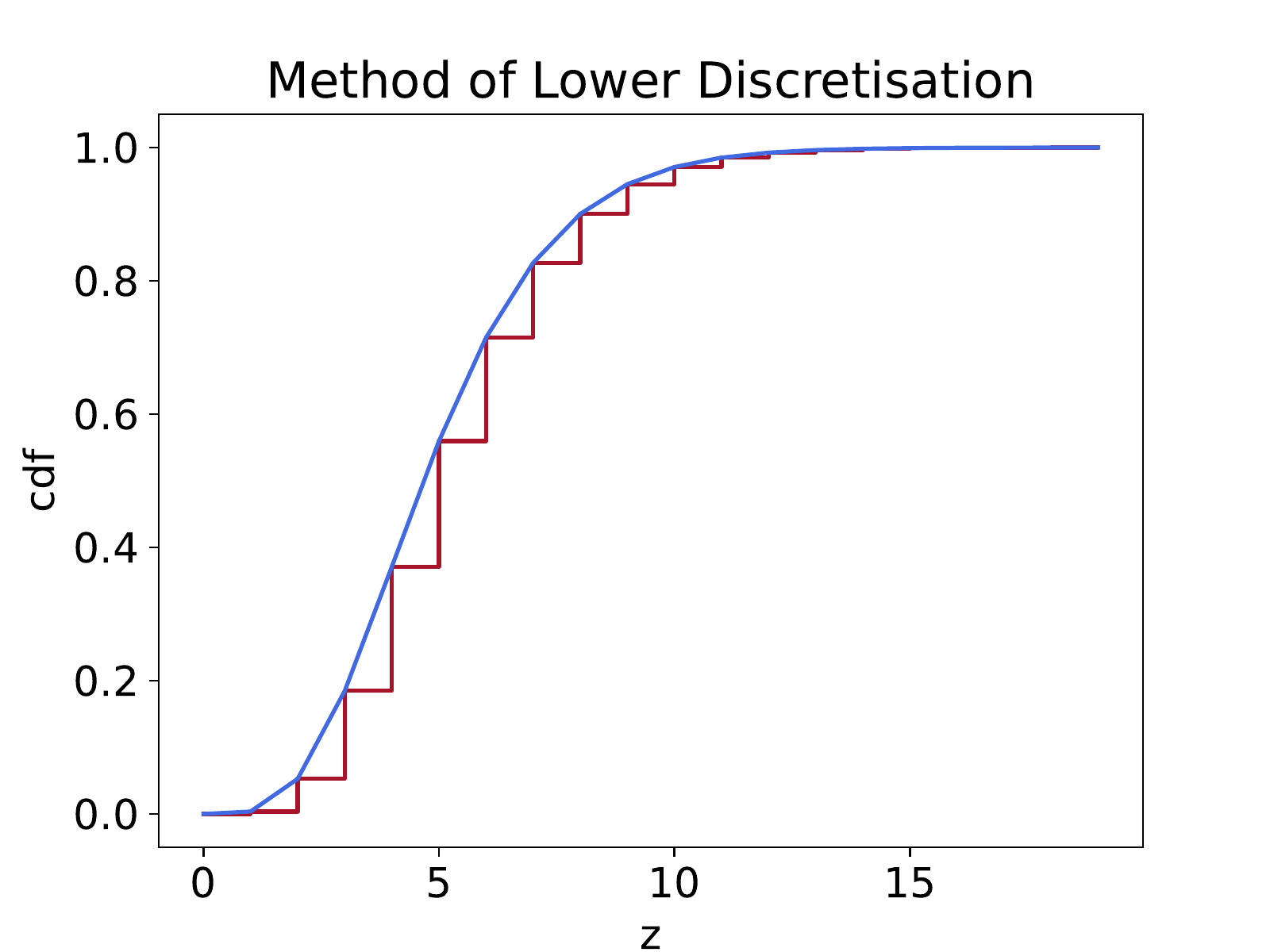}
\end{subfigure}
\hfill
\begin{subfigure}[b]{0.45\textwidth}
    \centering\includegraphics[width=8cm]{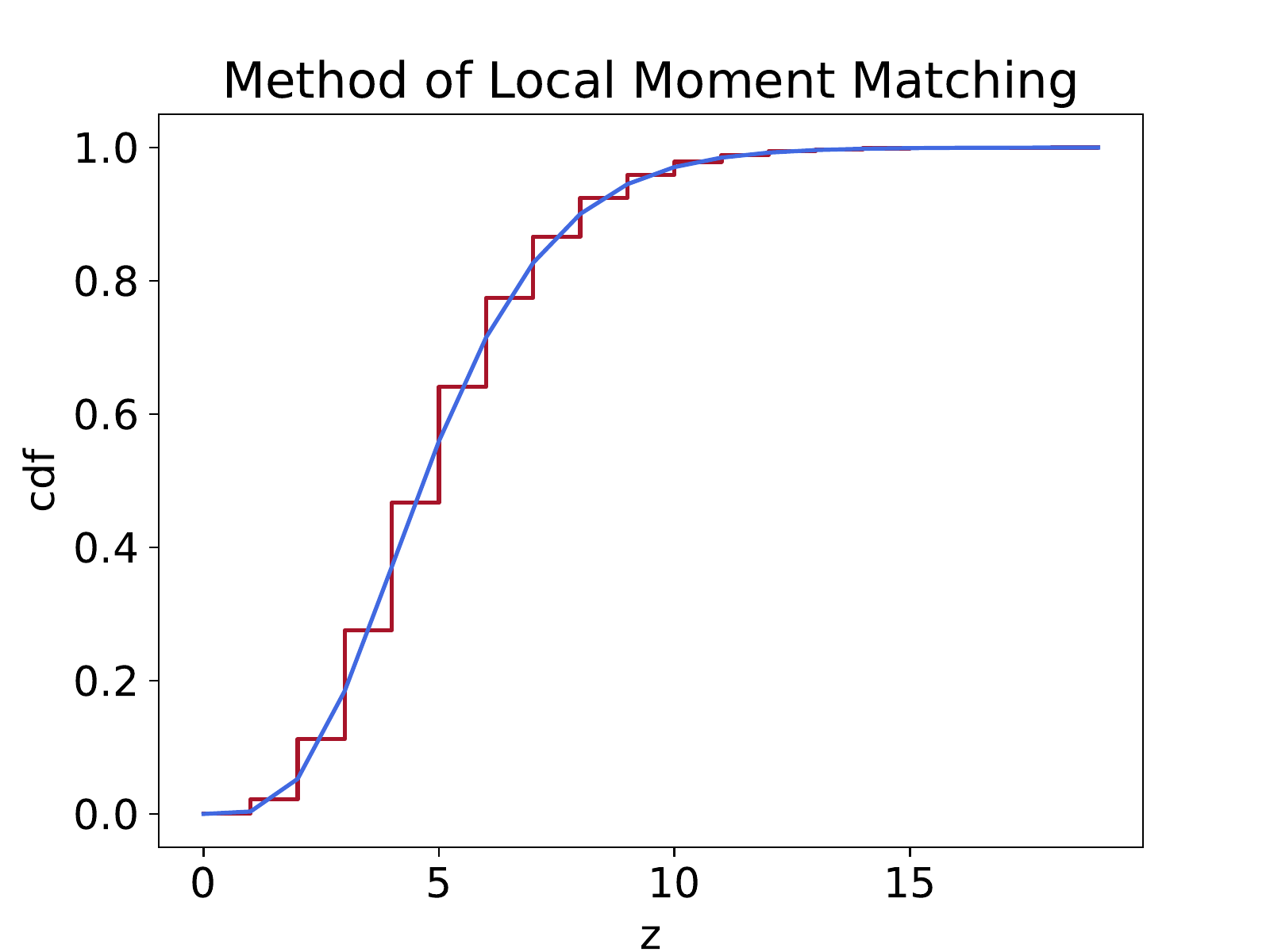}
\end{subfigure}
\caption{Illustration of the discretisation methods applied to a gamma($\text{\texttt{a}}=5$) severity. The graphs compare the original cdf (blue line) and the discretised (red line) cdf for mass dispersal (top left), upper discretisation (top right), lower discretisation (bottom left), and local moment matching (bottom right) methods. No coverage modifiers are present.}
\label{fig:discrmethods}
\end{figure}

\Cref{fig:discrmethods} illustrates and graphically contrasts the four different discretisation techniques.

The default discretisation method in \textbf{gemact} is the method of the mass dispersal, in which each $h j$ point is assigned with the probability mass of the $h$-span interval containing it (\Cref{fig:discrmethods} top left).
The upper discretisation and lower discretisation methods generate, respectively, pointwise upper and lower bounds to the true cdf. Hence, these can be used to provide a range where the original $F_z$ is contained (see \Cref{fig:discrmethods}, top right and bottom left graphs).
The method of local moment matching (\Cref{fig:discrmethods} bottom right) allows the moments of the original distribution $F_z$ to be preserved in the arithmetic distribution. A more general definition of this approach can be found in \citeA{gerber82}. We limited this approach to the first moment as, for higher moments, it is not well defined and it can potentially lead to a negative probability mass on certain lattice points \cite{embrechts09}.
The method of local moment matching shall be preferred for large bandwidths.
However, considering that there is no established analytical procedure to determine the optimal step, we remark that the choice of the discretisation step is an educated guess and it really depends on the problem at hand. In general, $h$ should be chosen such that it is neither too small nor too large relative to the severity losses. In the first case, the $hj$ points are not sufficient to capture the full range of the loss amount and the probability in the tail of the distribution exceeding the last discretisation node $h(m-1)$ is too large.
In the second case, the granularity of the severity distribution is not sufficient, and small losses are over-approximated. Additional rules of thumb and guidelines for the choice of discretisation parameters can be found in  \citeA[p. 248]{parodi14}. For example, one option is to perform the calculation with decreasing values of $h$ and check, graphically or according to a predefined criterion, whether the aggregate distribution changes substantially \cite{embrechts09}.
The reader should refer to \citeA[p. 179]{klugman98} and \citeA{embrechts09} for a more detailed treatment and discussion of discretisation methods. 

The above-mentioned discretisation methods are modified accordingly to reflect the cases where the transformation of \Cref{eq:minmax} is applied to the severity \cite[p.~517]{klugman98}. Below, we illustrate how to perform severity discretisation in \textbf{gemact}.

Once a continuous distribution is selected from those supported in our package (see \Cref{app:alldists}), the severity distribution is defined via the \texttt{Severity} class. 

\begin{minted}[mathescape]{python}
>>> from gemact.lossmodel import Severity
>>> severity = Severity(dist='gamma', par={'a': 5})
\end{minted}

The \texttt{dist} argument contains the name of the distribution, and 
the \texttt{par} argument specifies, as a dictionary, the distribution parameters. In the latter, each item key-value pair represents a distribution parameter name and its value. Refer to \texttt{distributions} module for a list of the distribution names and their parameter specifications.

The \texttt{discretize} method of the \texttt{Severity} class produces the discrete severity probability sequence according to the approaches described above. Below, we provide an example for mass dispersal. 

\begin{minted}[mathescape]{python}
>>> massdispersal = severity.discretize(
    discr_method='massdispersal',
    n_discr_nodes=50000,
    discr_step=.01,
    deductible=0
    )
\end{minted}

In order to perform the discretisation the following arguments are needed: 

\begin{itemize}
     \item The chosen discretisation method via \texttt{discr\_method}.
    \item The number of nodes ($m$) set in the \texttt{n\_discr\_nodes} argument. 
    \item The severity discretisation step ($h$) is in the \texttt{discr\_step} argument. 
    \item If necessary, a \texttt{deductible} specifying where the discretisation begins. The default value is zero.
\end{itemize}

After the discretisation is achieved, the mean of the discretised distribution can be calculated.

\begin{minted}[mathescape]{python}
>>> import numpy as np
>>> discrete_mean = np.sum(massdispersal['nodes'] * massdispersal['fj'])
>>> print('Mean of the discretised distribution:', discrete_mean)
Mean of the discretised distribution: 5.000000000000079
\end{minted}

Additionally, the arithmetic distribution obtained via the severity discretisation can be visually examined using the \texttt{plot\_discretized\_severity\_cdf}. This method is based on the \texttt{pyplot} interface to \textbf{matplotlib} \cite{Hunter:2007}. Hence, \texttt{plot\_discretized\_severity\_cdf} can be used together with \texttt{pyplot} functions, and can receive \texttt{pyplot.plot} arguments to change its output. In the following code blocks we adopt the \texttt{plot\_discretized\_severity\_cdf} method in conjunction with the \texttt{plot} function from \texttt{matplotlib.pyplot} to compare the cdf of a gamma distribution with mean and variance equal to $5$, with the arithmetic distribution obtained with the method of mass dispersal above. 
We first import the gamma distribution from the \texttt{distributions} module and compute the true cdf.

\begin{minted}[mathescape]{python}
>>> from gemact import distributions
>>> dist = distributions.Gamma(a=5)
>>> nodes = np.arange(0, 20)
>>> true_cdf = dist.cdf(nodes)
\end{minted}

Next, we plot the discrete severity using the \texttt{plot\_discr\_sev\_cdf} method. 

\begin{minted}[mathescape]{python}
>>> import matplotlib.pyplot as plt
>>> severity.plot_discr_sev_cdf(
    discr_method='massdispersal',
    n_discr_nodes=20,
    discr_step=1,
    deductible=0,
    color='#a71429'
    )
>>> plt.plot(nodes, true_cdf, color='#4169E1')
>>> plt.title('Method of Mass Dispersal')
>>> plt.xlabel('z')
>>> plt.show()
\end{minted}

The arguments \texttt{discr\_method}, \texttt{n\_discr\_nodes}, \texttt{discr\_step} and \texttt{deductible} can be used in the same manner as those described in the \texttt{discretize} method. The argument \texttt{color} is from the \texttt{matplotlib.pyplot.plot} method. The methods \texttt{title} and \texttt{xlabel} were also exported from \texttt{matplotlib.pyplot.plot} to add custom labels for the title and the x-axis \cite{Hunter:2007}.

The output of the previous code is shown in the top left graph of \Cref{fig:discrmethods}. To obtain the other graphs, simply set \texttt{discr\_method} to the desired approach, i.e. \texttt{`upper\_discretisation'}, \texttt{`lower\_discretisation'} or \texttt{`localmoments'}.

\subsection{Supported distributions}

The \textbf{gemact} package makes for the first time the $(a, b, 0)$ and $(a, b, 1)$ distribution classes \cite[p.~81]{klugman98} available in \textbf{Python}.
In the following code block we show how to use our implementation of the zero-truncated Poisson from the \texttt{distributions} module.

\begin{minted}[mathescape]{python}
>>> ztpois = distributions.ZTPoisson(mu=2)
\end{minted}

Each distribution supported in \textbf{gemact} has various methods and can be used in a similar fashion to any \textbf{scipy} distribution.
Next, we show how to compute the approximated mean via Monte Carlo simulation, with the random generator method for the \texttt{ZTPoisson} class.

\begin{minted}[mathescape]{python}
>>> random_variates = ztpois.rvs(10**5, random_state=1)
>>> print('Simulated Mean: ', np.mean(random_variates))
Simulated Mean: 2.3095
>>> print('Exact Mean: ', ztpois.mean())
Exact Mean: 2.3130352854993315
\end{minted}

Furthermore, supported copula functions can be accessed via the \texttt{copulas} module. Below, we compute the cdf of a two-dimensional Gumbel copula.

\begin{minted}[mathescape]{python}
>>> from gemact import copulas
>>> gumbel_copula = copulas.GumbelCopula(par=1.2, dim=2)
>>> values = np.array([[.5, .5]])
>>> print('Gumbel copula cdf: ', gumbel_copula.cdf(values)[0])
Gumbel copula cdf: 0.2908208406483879
\end{minted}

In the above example, it is noted that the copula parameter and dimension are defined by means of the \texttt{par} and the \texttt{dim} arguments, respectively.
The argument of the \texttt{cdf} method must be a \textbf{numpy} array whose dimensions meet the following requirements. Its first component is the number of points where the function shall be evaluated, its second component equals the copula dimension (\texttt{values.shape} of the example is in fact \texttt{(1,2)}).

The complete list of the distributions and copulas supported by \textbf{gemact} is available in \Cref{app:alldists}.
We remark that the implementation of some distributions is available in both \textbf{gemact} and \textbf{scipy.stats}. However, the objects of the \texttt{distributions} module include additional methods that are specific to their use in actuarial science. Examples are \texttt{lev} and \texttt{censored\_moment} methods, which allow the calculation of the limited expected value and censored moments of continuous distributions.
Furthermore, the choice of providing a \texttt{Severity} class is in order to have a dedicated object that includes functionalities relevant only for the calculation of a loss model, and not for distribution modelling in general. An example of this is the \texttt{discretize} method. A similar reasoning applies to the \texttt{Frequency} class.

\subsection{Illustration \texttt{lossmodel}}

\begin{figure}
    \centering
    \begin{tikzpicture}

    \draw (0,0) -- (3,0) -- (3,1.5) -- (0,1.5) -- (0,0);
    
    \draw (3.5+0,0) -- (3.5+3,0) -- (3.5+3,1.5) -- (3.5+0,1.5) -- (3.5+0,0);

    \draw (3.5+0,-2+0) -- (3.5+3,-2+0) -- (3.5+3,-2+1.5) -- (3.5+0,1.5-2) -- (3.5+0,0-2);
    
    \draw (7+0,0) -- (7+3,0) -- (7+3,1.5) -- (7+0,1.5) -- (7+0,0);

    \draw (7+0,2+0) -- (7+3,2+0) -- (7+3,2+1.5) -- (7+0,2+1.5) -- (7+0,2+0);
    
    \node[align=center] at (1.5,.75) 
    {\texttt{Frequency}};
    \node[align=center] at (3.5+1.5,.75) 
    {\texttt{Severity}};
    \node[align=center] at (3.5+1.5,.75-2) 
    {\texttt{LossModel}};
    \node[align=center] at (7+1.5,.75) 
    {\texttt{PolicyStructure}};
    \node[align=center] at (7+1.5,2+.75) 
    {\texttt{Layer}};

    \draw[->]  (3.5+1.5,0) -- (3.5+1.5,-.4) ;
    \draw[->]  (7+1.5,2) -- (7+1.5,1.6) ;
    
    \draw[->]  (1.5,-.5-.75) -- (3.4,-.5-.75) ;
    \draw[->]  (7+1.5,-.5-.75)  -- (6.6,-.5-.75)  ;
    \draw  (7+1.5,0) -- (7+1.5,-.5-.75) ;
    \draw  (1.5,0) -- (1.5,-.5-.75) ;
    
\end{tikzpicture}

    \caption{Class diagram of the \texttt{lossmodel} module. A rectangle represents a class; an arrow connecting two classes indicates that the target class employs the origin class as an attribute. In this case, a \texttt{LossModel} object entails  \texttt{Frequency}, \texttt{Severity} and \texttt{PolicyStructure} class instances. These correspond to the frequency model, the severity model and the policy structure, respectively. The latter, in particular, is in turn specified via one or more \texttt{Layer} objects, which include coverage modifiers of each separate policy component.}
    \label{fig:lossmodeldiagram}
\end{figure}
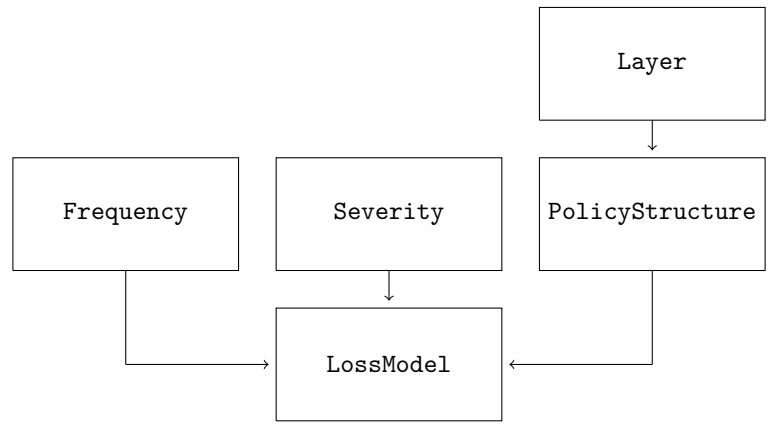

The following are examples of how to get started and use \texttt{lossmodel} module and its classes for costing purposes. As an overview, \Cref{fig:lossmodeldiagram} schematises the class diagram of the \texttt{lossmodel} module, highlighting its structure and the dependencies of the \texttt{LossModel} class.

The \texttt{Frequency} and the \texttt{Severity} classes represent, respectively, the frequency and the severity components of a loss model.
In these, \texttt{dist} identifies the name of the distribution, and 
\texttt{par} specifies its parameters as a dictionary, in which each item key-value pair corresponds to a distribution parameter name and value. Please refer to \texttt{distributions} module for the full list of the distribution names and their parameter specifications.
The code block below shows how to initiate a frequency model.

\begin{minted}[mathescape]{python}
>>> from gemact.lossmodel import Frequency
>>> frequency = Frequency(
    dist='poisson',
    par={'mu': 4},
    threshold=0
    )
\end{minted}

In practice, losses are reported only above a certain threshold (the reporting threshold) and the frequency model can be estimated only above another, higher threshold
called the analysis threshold \cite[p. 323]{parodi14}. This can be specified in \texttt{Frequency} with the optional parameter \texttt{threshold}, whose default value is 0 (i.e. the analysis threshold equals the reporting threshold).
Severity models in \textbf{gemact} always refer to the reporting threshold.

A loss model is defined and computed through the \texttt{LossModel} class. Specifically, \texttt{Frequency} and \texttt{Severity} objects are assigned to \texttt{frequency} and \texttt{severity} arguments of \texttt{LossModel} to set the parametric assumptions of the frequency and the severity components. Below, we use the \texttt{severity} object we instanced in \Cref{sec:severitydiscretisation}.

\begin{minted}[mathescape]{python}
>>> from gemact.lossmodel import LossModel
>>> lm_mc = LossModel(
    frequency=frequency,
    severity=severity,
    aggr_loss_dist_method='mc',
    n_sim=10**5,
    random_state=1
    )
INFO:lossmodel|Approximating aggregate loss distribution via Monte Carlo simulation
INFO:lossmodel|MC simulation completed
\end{minted}

In the previous example, in more detail, \texttt{lm\_mc} object adopts the Monte Carlo simulation for the calculation of the aggregate loss distribution. This approach is set via the \texttt{aggr\_loss\_dist\_method} equal to \texttt{`mc'}.
The additional parameters required for the simulation are: 

\begin{itemize}
    \item the number of simulations \texttt{n\_sim},
    \item the (pseudo)random number generator initialiser \texttt{random\_state}.
\end{itemize}

The cdf of the aggregate loss distribution can be displayed with the \texttt{plot\_dist\_cdf} method.
Moreover, a recap of the computation specifications can be printed with the \texttt{print\_aggr\_loss\_specs} method.

\begin{minted}[mathescape]{python}
>>> lm_mc.print_aggr_loss_method_specs()
Aggregate Loss Distribution: layer 1
======================================================
                        Quantity           Value
======================================================
       Aggregate loss dist. method              mc
              Number of simulation          100000
                      Random state               1
\end{minted}

The aggregate loss mean, standard deviation and skewness can be accessed with the \texttt{mean}, \texttt{std}, and \texttt{skewness} methods, respectively. The code below shows how to use these methods.
\begin{minted}[mathescape]{python}
>>> lm_mc.mean(use_dist=True)
19.963155575580227
>>> lm_mc.mean(use_dist=False)
20.0
>>> lm_mc.coeff_variation(use_dist=True)
0.5496773171375182
>>> lm_mc.coeff_variation(use_dist=False)
0.5477225575051661
>>> lm_mc.skewness(use_dist=True)
0.6410913579225725
>>> lm_mc.skewness(use_dist=False)
0.6390096504226938
\end{minted}
When the \texttt{use\_dist} argument is set to \texttt{True} (default value) the quantity is derived from the approximated aggregate loss distribution (\texttt{dist} property). Conversely, when it is \texttt{False}, the calculation relies on the closed-form formulas of the moments of the aggregate loss random variable.  These can be obtained directly from the closed-form moments of the frequency and the severity model transformed according to the coverage modifiers \cite[p. 322]{parodi14}.
This option is available for \texttt{mean}, \texttt{std}, \texttt{var} (i.e. the variance), \texttt{coeff\_variation} (i.e. the coefficient of variation) and \texttt{skewness} methods.
It should be noted that the calculation with \texttt{use\_dist=False} is not viable when aggregate coverage modifiers are present. In such circumstance, the method must necessarily be based on the approximated aggregate loss distribution.
In any situations, it is possible to get the moments of the approximated aggregate loss distribution via the \texttt{moment} method. The \texttt{central} and \texttt{n} arguments specify, respectively, whether the moment is central and the order of the moment.
\begin{minted}[mathescape]{python}
>>> lm_mc.moment(central=False, n=1)
19.963155575580227
\end{minted}
Furthermore, for the aggregate loss distribution, the user can simulate random variates via the \texttt{rvs} method. The quantile and the cdf functions can be computed via the \texttt{ppf} and the \texttt{cdf} methods.
Below is an example of the \texttt{ppf} method returning the 0.80- and 0.70-level quantiles.

\begin{minted}[mathescape]{python}
>>> lm_mc.ppf(q=[0.80, 0.70])
array([28.81343738, 24.85497983])
\end{minted}

The following code block shows the costing of a `20 excess 5' XL reinsurance contract. Coverage modifiers are set in a \texttt{PolicyStructure} object.

\begin{minted}[mathescape]{python}
from gemact.lossmodel import PolicyStructure, Layer
>>> policystructure = PolicyStructure(
    layers=Layer(
    cover=20,
    deductible=5
    ))
\end{minted}

More precisely, in the \texttt{Layer} class the contract \texttt{cover} and the \texttt{deductible} are provided.
Once the model assumptions are set and the policy structure is specified, the aggregate loss distribution can be computed.

\begin{minted}[mathescape]{python}
>>> lm_XL = LossModel(
    frequency=frequency,
    severity=severity,
    policystructure=policystructure,
    aggr_loss_dist_method='fft',
    sev_discr_method='massdispersal',
    n_aggr_dist_nodes=2**17
    )
INFO:lossmodel|Approximating aggregate loss distribution via FFT
INFO:lossmodel|FFT completed
\end{minted}

It can be noted that, in the previous code block, we determined the aggregate loss distribution with \texttt{`fft'} as \texttt{aggr\_loss\_dist\_method}. In such case, \texttt{LossModel} requires additional arguments for defining the computation process. Namely:
\begin{itemize}
    \item The number of nodes of the aggregate loss distribution \texttt{n\_aggr\_dist\_nodes}.
    \item The method to discretise the severity distribution \texttt{sev\_discr\_method}. Above, we opted for the method of mass dispersal (\texttt{`massdispersal'}). 
    \item The number of nodes of the discretised severity \texttt{n\_sev\_discr\_nodes} (optional).
    \item The discretisation step \texttt{sev\_discr\_step} (optional). When a cover is present, \textbf{gemact} automatically adjusts the discretisation step parameter to have the correct number of nodes in the transformed severity support.
\end{itemize}
The same arguments shall be specified when computing the aggregate loss distribution with the recursive formula, i.e. \texttt{aggr\_loss\_dist\_method} set to \texttt{`recursive'}.

The costing specifications of a \texttt{LossModel} object can be accessed with the method \texttt{print\_costing\_specs()}.

\begin{minted}[mathescape]{python}
>>> lm_XL.print_costing_specs()
                               Costing Summary: Layer 1 
==========================================================================
                                               Quantity           Value
==========================================================================
                                                  Cover            20.0
                                             Deductible             5.0
                                        Aggregate cover             inf
                                   Aggregate deductible               0
        Pure premium (dist est.) before share partecip.            3.51
                    Pure premium before share partecip.            3.51
                                        Share partecip.               1
                               Pure premium (dist est.)            3.51
                                           Pure premium            3.51
\end{minted}

The previous output exhibits a summary of the contract structure (cover, deductible, aggregate cover, aggregate deductible) and details about the costing results. It is noted that the default value of the share participation equals 1, i.e. $a$ is equal to $1$ in \Cref{eq:qs}.

Similar to the previous example, user can access moments of the aggregate loss calculated from the approximated distribution and from the closed-from solution.

\begin{minted}[mathescape]{python}
>>> lm_XL.mean(use_dist=True)
3.509346100359707
>>> lm_XL.mean(use_dist=False)
3.50934614394912
>>> lm_XL.coeff_variation(use_dist=True)
1.0001481880266856
>>> lm_XL.coeff_variation(use_dist=False)
1.0001481667319252
>>> lm_XL.skewness(use_dist=True)
1.3814094240544392
>>> lm_XL.skewness(use_dist=False)
1.3814094309741256
\end{minted}

The next example illustrates the costing of an XL with reinstatements. The \texttt{PolicyStructure} object is set as follows.
\begin{minted}[mathescape]{python}
>>> policystructure_RS = PolicyStructure(
    layers=Layer(
    cover=100,
    deductible=0,
    aggr_deductible=100,
    reinst_percentage=1,
    n_reinst=2
    ))
\end{minted}
The relevant parameters are:
\begin{itemize}
    \item the aggregate deductible parameter \texttt{aggr\_deductible},
    \item the number of reinstatements \texttt{n\_reinst},
    \item  the reinstatement percentage \texttt{reinst\_percentage}.
\end{itemize}
Below, we compute the pure premium of \Cref{eq:reinstatementsP}, given the parametric assumptions on the frequency and the severity of the loss model.

\begin{minted}[mathescape]{python}
>>> lm_RS = LossModel(
    frequency=Frequency(
    dist='poisson',
    par={'mu': .5}
    ),
    severity=Severity(
    dist='pareto2',
    par={'scale': 100, 'shape': 1.2}
    ),
    policystructure = policystructure_RS,
    aggr_loss_dist_method='fft',
    sev_discr_method='massdispersal',
    n_aggr_dist_nodes=2**17
    )
>>> print('Pure premium (RS): ', lm_RS.pure_premium_dist[0])
Pure premium (RS): 4.319350355177216
\end{minted}

A \texttt{PolicyStructure} object can handle multiple layers simultaneously. These can overlap and do not need to be contiguous. The \texttt{length} property indicates the number of layers. The code block below deals with three \texttt{Layer} objects, the first without aggregate coverage modifiers and (participation) share of 0.5, the second with reinstatements, the third with an aggregate cover.

\begin{minted}[mathescape]{python}
>>> policystructure=PolicyStructure(
    layers=[
    Layer(cover=100, deductible=100, share=0.5),
    Layer(cover=200, deductible=100, n_reinst=2, reinst_percentage=0.6),
    Layer(cover=100, deductible=100, aggr_cover=200)
    ])
>>> lossmodel_multiple = LossModel(
    frequency=Frequency(
    dist='poisson',
    par={'mu': .5}
    ),
    severity=Severity(
    dist='genpareto',
    par={'loc': 0, 'scale': 83.34, 'c': 0.834}
    ),
    policystructure=policystructure
    )
WARNING:lossmodel|Aggregate loss distribution calculation is omitted as
                    aggr_loss_dist_method is missing
WARNING:lossmodel|Layer 2: costing is omitted as aggr_loss_dist_method is missing    
WARNING:lossmodel|Layer 3: costing is omitted as aggr_loss_dist_method is missing
\end{minted}
As outlined by the warning messages, since the instantiation of \texttt{lossmodel\_multiple} lacks of \texttt{aggr\_loss\_dist\_method}, the calculation of the aggregate loss distribution is omitted. Therefore, costing results are accessible solely for the first \texttt{Layer}, since this is the only one without aggregate coverage modifiers. This fact is reflected in \texttt{pure\_premium} and \texttt{pure\_premium\_dist} properties, containing the premiums derived from the closed-form means and the approximated aggregate loss distribution means, respectively. The latter are indeed not available. 
\begin{minted}[mathescape]{python}
>>> lossmodel_multiple.pure_premium
[8.479087307840043, None, None]
>>> lossmodel_multiple.pure_premium_dist
[None, None, None]
\end{minted}
Contrarily, once the aggregate loss distribution is determined (via \texttt{dist\_calculate} method), all layer premiums in \texttt{pure\_premium\_dist} are available from the costing (\texttt{costing} method). As expected, \texttt{pure\_premium} content remains unaffected.
\begin{minted}[mathescape]{python}
>>> lossmodel_multiple.dist_calculate(
    aggr_loss_dist_method='fft',
    sev_discr_method='massdispersal',
    n_aggr_dist_nodes=2**17
    )
INFO:lossmodel|Computation of layers started
INFO:lossmodel|Computing layer: 1
INFO:lossmodel|Approximating aggregate loss distribution via FFT
INFO:lossmodel|FFT completed
INFO:lossmodel|Computing layer: 2
INFO:lossmodel|Approximating aggregate loss distribution via FFT
INFO:lossmodel|FFT completed
INFO:lossmodel|Computing layer: 3
INFO:lossmodel|Approximating aggregate loss distribution via FFT
INFO:lossmodel|FFT completed
INFO:lossmodel|Computation of layers completed
>>> lossmodel_multiple.costing()
>>> lossmodel_multiple.pure_premium_dist
[8.479087307062226, 25.99131088702302, 16.88704720494799]
>>> lossmodel_multiple.pure_premium
[8.479087307840043, None, None]
\end{minted}
At last, it is remarked that each \texttt{Layer} in \texttt{PolicyStructure} is associated with an index \texttt{idx}, starting from 0, based on the layer order of the instantiation of the \texttt{PolicyStructure} object. This is of help when the user needs to retrieve particular information and features, or to apply methods to one specific layer. All the methods that include the \texttt{idx} argument have 0 as default value, meaning that they are applied to the first (or only) layer unless otherwise specified. For example, the \texttt{print\_policy\_layer\_specs} method produces a table of recap of the features of the layer indicated by the \texttt{idx} argument. Below  \texttt{idx} equals 1, namely the second Layer in \texttt{policystructure}.
\begin{minted}[mathescape]{python}
>>> lossmodel_multiple.print_policy_layer_specs(idx=1)
      Policy Structure Summary: layer 2 
==================================================
              Specification           Value
==================================================
                     Deductible           100.0
                          Cover           200.0
           Aggregate deductible               0
           Reinstatements (no.)               2
     Reinst. layer percentage 1             0.6
     Reinst. layer percentage 2             0.6
            Share partecipation               1
\end{minted}
Likewise, the code block below returns the aggregate loss distribution mean of the third \texttt{Layer}, as \texttt{idx} is set to 2.
\begin{minted}[mathescape]{python}
>>> lossmodel_multiple.mean(idx=2)
16.88704720494799
\end{minted}

\subsection{Comparison of the methods for computing the aggregate loss distribution}
\label{ss:computationalmethodscomparison}

In this section we analyse accuracy and speed of the computation of the aggregate loss distribution using fast Fourier transform (FFT), recursive formula (recursion), and Monte Carlo simulation (MC) approaches, as the number of nodes, the discretisation step and the number of simulations vary.

For this purpose, a costing example was chosen such that the analytical solutions of the moments of the aggregate loss distributions are known and can be compared to those obtained from the approximated aggregate loss distribution. The values of the parameters of the severity and frequency models are taken from the illustration in \citeA[p.~262]{parodi14}. Specifically, these and the policy structure specifications are as follows.

\begin{itemize}
    \item Severity: lognormal distribution with parameters shape $=1.3$ and  scale $=36315.49$, hence whose mean and standard deviation are $84541.68$ and $177728.30$, respectively.
    \item Frequency: Poisson distribution with parameter $\mu = 3$, with analysis threshold $d$. It belongs to the $(a, b, 0)$ family described in \Cref{ss:recursiveformula} with parameters $a=0$, $b=3$ and $p_0=e^{-3}$.
    \item Policy structure: contract with deductible $d=10000$.
\end{itemize}

In particular, the accuracy in the approximation of the aggregate loss distribution has been assessed using the relative error in the estimate of the mean, coefficient of variation (CoV) and skewness, with respect to their reference values (i.e. error = estimate/reference - 1). The latter are obtained using the following closed-form expressions \cite[p.~382]{Bean2000ProbabilityTS}. 
\begin{align*}
    & \mathbb{E}\left[X\right] = \mu \mathbb{E}\left[ L_{d, \infty} (Z) \right] ,\\
    & \text{CoV}\left[X\right] =  \dfrac{ \mathbb{E}\left[ L_{d, \infty} (Z)^2 \right]^{1/2}}{\mu^{1/2} \mathbb{E}\left[ L_{d, \infty} (Z) \right]} ,\\
    & \text{Skewness}\left[X\right] =  \dfrac{ \mathbb{E}\left[ L_{d, \infty} (Z)^3 \right]}{\mu^{1/2} \mathbb{E}\left[ L_{d, \infty} (Z)^2 \right]^{3/2}}.
\end{align*}

Their values for the example are in the top part of \Cref{tab:lossmodel_comp_accuracy}.
The test of the speed of our implementation has been carried out by measuring the execution time of the approximation of the aggregate loss distribution function with the built-in \texttt{timeit} library. In line with best practice \cite[Chapter 18]{martelli05}, the observed minimum execution time of independent repetitions of the function call was adopted. 

\begin{table}[!htbp]
\centering
{
\begin{tabular}{l|r|r|r}
  \toprule
 \multicolumn{1}{c}{ } & \multicolumn{1}{c}{Mean}  & \multicolumn{1}{c}{CoV} & \multicolumn{1}{c}{Skewness} \\
 \midrule
 Reference Values  & 268837 &  1.35520 &  7.02399 \\
\bottomrule 
\end{tabular}
}

\vspace{2mm}
\centering
\begin{tabular}{l|r|r|r|r}
\toprule
 \multicolumn{1}{c}{Method} & \multicolumn{1}{c}{Time (sec.)}& \multicolumn{1}{c}{Mean}  & \multicolumn{1}{c}{CoV} & \multicolumn{1}{c}{Skewness} \\
\midrule
FFT ($h = 50, m = 2^{**}14$)& 0.002& -2.76451e-01& -3.00646e-01& -8.06071e-01\\
FFT ($h = 100, m = 2^{**}14$)& 0.002& -8.96804e-02& -1.98006e-01& -7.15193e-01\\
FFT ($h = 200, m = 2^{**}14$)& 0.003& -2.19140e-02& -1.01312e-01& -5.69628e-01\\
FFT ($h = 400, m = 2^{**}14$)& 0.003& -4.41036e-03& -5.88141e-02& -3.47025e-01\\
FFT ($h = 50, m = 2^{**}16$)& 0.010&  -2.19153e-02& -1.01317e-01& -5.69635e-01\\
FFT ($h = 100, m = 2^{**}16$)& 0.009& -4.40959e-03& -5.88200e-02& -3.47044e-01\\
FFT ($h = 200, m = 2^{**}16$)& 0.008& -7.24988e-04& -1.36155e-02& -2.37709e-01\\
FFT ($h = 400, m = 2^{**}16$)& 0.007& -9.66263e-05& -3.49930e-03& -1.14264e-01\\
FFT ($h = 50, m = 2^{**}18$)& 0.037& -7.24464e-04& -1.36142e-02& -2.37694e-01\\
FFT ($h = 100, m = 2^{**}18$)& 0.035& -9.34546e-05& -3.46164e-03& -1.13558e-01\\
FFT ($h = 200, m = 2^{**}18$)& 0.036& -3.61429e-06& -3.69322e-04& -3.07951e-02\\
FFT ($h = 400, m = 2^{**}18$)& 0.041& -1.73491e-06& -2.80936e-05& -6.37228e-03\\
\midrule
Recursion ($h = 50, m = 2^{**}14$)& 1.397& -2.76451e-01& -3.00646e-01& -8.06071e-01\\
Recursion ($h = 100, m = 2^{**}14$)& 1.368& -8.96804e-02& -1.98006e-01& -7.15193e-01\\
Recursion ($h = 200, m = 2^{**}14$)& 1.398& -2.19140e-02& -1.01312e-01& -5.69628e-01\\
Recursion ($h = 400, m = 2^{**}14$)& 1.390& -4.41046e-03& -4.16016e-02& -4.00405e-01\\
Recursion ($h = 50, m = 2^{**}16$)& 13.401& -2.19154e-02& -1.01317e-01& -5.69636e-01\\
Recursion ($h = 100, m = 2^{**}16$)& 12.606& -4.41006e-03& -4.16096e-02& -4.00430e-01\\
Recursion ($h = 200, m = 2^{**}16$)& 11.815& -7.25738e-04& -1.36242e-02& -2.37782e-01\\
Recursion ($h = 400, m = 2^{**}16$)& 11.771& -9.62811e-05& -3.49125e-03& -1.14125e-01\\
Recursion ($h = 50, m = 2^{**}18$)& 448.168& -7.25465e-04& -1.36260e-02& -2.37794e-01\\
Recursion ($h = 100, m = 2^{**}18$)& 464.185& -9.47567e-05& -3.49381e-03& -1.14141e-01\\
Recursion ($h = 200, m = 2^{**}18$)& 460.200& -1.00139e-05& -6.92170e-04& -4.30010e-02\\
Recursion ($h = 400, m = 2^{**}18$)& 462.339& -2.41068e-06& -1.03582e-04& -1.25319e-02\\
\midrule
MC ($2^{**}14$ sim.)& 0.243& -1.37610e-02& -3.67078e-02& -3.77663e-01\\
MC ($2^{**}16$ sim.)& 0.985& -5.53096e-03& -2.44089e-02& -3.49520e-01\\
MC ($2^{**}18$ sim.)& 3.922& -3.02675e-03& -6.31524e-03& -8.93550e-02\\
MC ($2^{**}20$ sim.)& 15.901&  -1.20257e-03& -5.21299e-03& -7.49223e-02\\
\bottomrule
\end{tabular}
\caption{\label{tab:lossmodel_comp_accuracy} Accuracy and speed of the approximation of the aggregate loss distribution using fast Fourier transform (FFT), the recursive formula (recursion), and the Monte Carlo simulation (MC) when varying the number of nodes ($m$), the discretisation step ($h$) and number of simulations. The upper table contains the reference values obtained from the closed-form solutions. The lower table reports the execution times in second and the relative errors with respect to the reference values.
}
\end{table}

The results of the analysis are reported in \Cref{tab:lossmodel_comp_accuracy}. We considered for FFT and recursion the values 50, 100, 200 and 400 for the discretisation step $h$, and $2^{14}, 2^{16}, 2^{18}$ for the number of nodes. It can be noted that FFT and recursion produce similar figures in terms of accuracy but the former is drastically faster. This is expected as FFT takes essentially $\mathcal{O}(m \log(m))$ operations, compared to the $\mathcal{O}(m^2)$ operations for recursion \cite{embrechts09}.
Furthermore, for this example, in both FFT and recursion, when $h$ increases, the error reduces.
Finally, MC approach lies in between the two other alternatives, when it comes to computing time.

\subsection{Comparison with \textbf{aggregate} FFT implementation}
\label{ss:aggregatecomparison}

The \textbf{aggregate} package \cite{aggregatepackage} allows to compute the aggregate loss distribution using FFT. In this section, we compare our implementation with \textbf{aggregate} implementation to show that the two provide similar results.
We adopt the same underlying frequency and severity assumptions of \Cref{ss:computationalmethodscomparison}, and contracts with different combinations of individual and aggregate coverage modifiers. In particular, we first consider no reinsurance, then an Excess-of-loss (XL), with individual-only coverage modifiers, a Stop Loss (SL) and finally an Excess-of-loss with individual and aggregate coverage modifiers (XL w/agg.). In line with the example in \citeA[p.~262]{parodi14}, individual coverage modifiers are $c= 1 000 000$ and $d= 10 000$, aggregate coverage modifiers are $u= 1 000 000$ and $v= 50 000$.
The number of nodes $m$ is set to $2^{22}$ in all the calculations. 

The comparison of the speed of the two implementations has been carried out by means of the built-in \texttt{timeit} library. In particular we measured the execution time of both the initialisation of the main computational object and the calculation of the aggregate loss distribution, in order to make the comparison consistent and adequate. In line with best practice \cite[Chapter 18]{martelli05}, the observed minimum execution time of independent repetitions of the function call was adopted.

As reported by \Cref{tab:gemactvsaggregate}, the two implementations generate consistent results; their estimates for mean, CoV and skewness tend to coincide for all contracts and are close to the reference values when these are available. When it comes to computing time, \textbf{gemact} takes a similar time, just under one second, for all the examples considered. For the case without reinsurance and the XL, \textbf{aggregate} performs slightly better. Conversely, for SL and XL w/agg., \textbf{gemact} is more than twice as fast.

\begin{table}[ht]
\centering
\begin{tabular}{l|l|r|r|r|r}
  \toprule
  \multicolumn{1}{l}{Contract} &\multicolumn{1}{c}{Library} 
  &\multicolumn{1}{c}{Time (sec.)} &\multicolumn{1}{c}{Mean} & \multicolumn{1}{c}{CoV}& \multicolumn{1}{c}{Skewness} \\ 
  \midrule
  \multirow{3}{*}{No reinsurance} 
 & Reference value & - & 253625 &  1.34406 &  7.28410 \\
 & \textbf{gemact}  & 0.9581&  253625 &  1.34406 &  7.27745 \\
 & \textbf{aggregate} & 0.8102&  253625 &  1.34406 &  7.28346 \\
 \midrule
 \multirow{3}{*}{XL} 
 & Reference value & - & 256355 &  1.10772 &  2.08525 \\
 & \textbf{gemact} & 0.9384&  256355 &  1.10772 &  2.08525 \\
 & \textbf{aggregate} & 0.8445&  256354 &  1.10773 &  2.08527 \\
\midrule
\multirow{3}{*}{SL } 
 & Reference value & - & - & - & -\\
 & \textbf{gemact} & 0.9168&  194143 &  1.24438 &  1.73658 \\
 & \textbf{aggregate}  & 2.4254&  194143 &  1.24438 &  1.73658 \\
\midrule
\multirow{3}{*}{XL w/agg.} 
 & Reference value & - & - & - & -\\
& \textbf{gemact} & 0.9388&  206363 &  1.22464 &  1.62335 \\
& \textbf{aggregate}  & 2.5221&  206363 &  1.22465 &  1.62336 \\
   \bottomrule
\end{tabular}

\caption{\label{tab:gemactvsaggregate} Comparison for different contracts of \textbf{aggregate} and \textbf{gemact} implementation of the aggregate loss distribution computation via FFT. When there are no aggregate coverage modifiers, reference values are given in addition to estimated ones. For the XL and the XL w/agg. contracts, individual conditions are $c= 1 000 000$ and $d= 10 000$; for the SL and the XL w/agg. contracts, aggregate coverage modifiers are $u= 1 000 000$ and $v= 50 000$. Execution times are expressed in seconds.
}
\end{table}

\section{Loss aggregation}
\label{sec:lossaggregation}

In insurance and finance, the study of the sum of dependent random variables is a central topic. A notable example is risk management, where the distribution of the sum of certain risks needs to be approximated and analysed for solvency purposes \cite{Wilhelmy2010}.
Another application is the pricing of financial and (re)insurance contracts where the payout depends on the aggregation of two or more dependent outcomes \cite<see for example>{Cummins1999, Wang2013}. In this section, in contrast with the collective risk theory in \Cref{sec:lossmodel}, we model the sum of a given number of random variables $d > 1$ that are neither independent nor necessarily identically distributed.

More specifically, consider now the random vector

$$\left(X_1, \ldots, X_d\right): \Omega \rightarrow 
\mathbb{R}^d,$$ 

whose joint cdf 

\begin{equation}
\label{eq:jointcdf}
    H\left(x_1,\ldots,x_d\right)=P\left[X_1\le x_1,\ldots,X_d\le x_d\right]  
\end{equation} 

is known analytically or can be numerically evaluated in an efficient way. For a real threshold $s$, the \texttt{gemact} package implements the \textit{AEP algorithm} and a \textit{Monte Carlo simulation} approach to model 

\begin{equation}
\label{eq:simplexprob}
    P\left[X_1 + \ldots + X_d \leq s\right],
\end{equation}

given a set of parametric assumptions on the one-dimensional marginals $X_1, ..., X_d$ and their copula.

More specifically, the AEP algorithm is designed to approximate \Cref{eq:jointcdf} through a geometric procedure, without relying on simulations or numerical integration of a
density. In \Cref{app:AEP} a brief description of the algorithm is given. For a complete mathematical treatment of the subject, the reader should refer to \citeA{arbenz11}.

\subsection{Illustration \texttt{lossaggregation}}

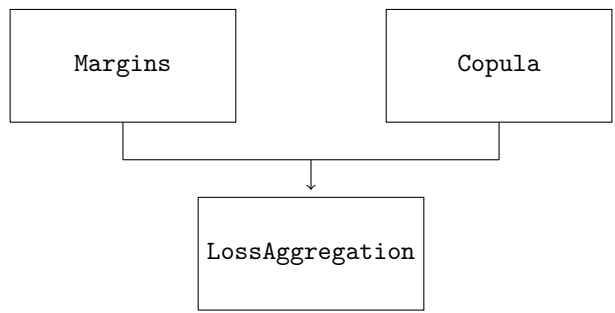
\begin{figure}
    \centering
    \begin{tikzpicture}

    \draw (0,0) -- (3,0) -- (3,1.5) -- (0,1.5) -- (0,0);
    
    \draw (5+0,0) -- (5+3,0) -- (5+3,1.5) -- (5+0,1.5) -- (5+0,0);

    \draw (2.5+0,-2.5+0) -- (2.5+3,-2.5+0) -- (2.5+3,-2.5+1.5) -- (2.5+0,1.5-2.5) -- (2.5+0,0-2.5);
    
    \node[align=center] at (1.5,.75) 
    {\texttt{Margins}};
    
    \node[align=center] at (5+1.5,.75) 
    {\texttt{Copula}};
    \node[align=center] at (2.5+1.5,.75-2.5) 
    {\texttt{LossAggregation}};

    \draw  (1.5,0) -- (1.5,-.5) ;
    \draw  (5+1.5,0) -- (5+1.5,-.5) ;
    \draw  (1.5,-.5)-- (5+1.5,-.5) ;
    \draw[->] (2.5+1.5,-.5) -- (2.5+1.5,-.9) ;
    
\end{tikzpicture}
    \caption{Class diagram of the \texttt{lossaggregation} module. A rectangle represents a class; an arrow connecting two classes indicates that the target class employs the origin class as an attribute.
    In this case, a \texttt{LossAggregation} object entails \texttt{Margins} and \texttt{Copula} class instances.}
    \label{fig:lossaggregationdiagram}
\end{figure}

Below are some examples of how to use the \texttt{LossAggregation} class. This belongs to the \texttt{lossaggregation} module, whose class diagram is depicted in \Cref{fig:lossaggregationdiagram}. The main class is \texttt{LossAggregation}, which is the computation object of the random variable sum. This depends on the classes \texttt{Margins} and \texttt{Copula}. Evidently, the former represents the marginal distributions and the latter describes the dependency structure, i.e. the copula.

Consistent with the \textbf{gemact} framework, the specifications needed to instantiate \texttt{Margins} and \texttt{Copula} are akin to those of the \texttt{Frequency} and \texttt{Severity} classes in \texttt{lossmodel}.
\texttt{Copula} objects are specified by:
\begin{itemize}
    \item \texttt{dist}: the copula distribution name as a string (\texttt{str}),
    \item \texttt{par}: the parameters of the copula, as a \texttt{dictionary}.
\end{itemize}
Likewise, \texttt{Margins} objects are defined by:
\begin{itemize}
    \item \texttt{dist}: the \texttt{list} of marginal distribution names as strings (\texttt{str}),
    \item \texttt{par}: the \texttt{list} of parameters of the marginal distributions, each \texttt{list} item is a \texttt{dictionary}.
\end{itemize}
Please refer to \Cref{tab:alldists} and \Cref{tab:allcopulas} of \Cref{app:alldists} for the complete list of the supported distributions and copulas.

\begin{minted}[mathescape]{python}
>>> from gemact import LossAggregation, Copula, Margins
>>> lossaggregation = LossAggregation(
    margins=Margins(
    dist=['genpareto', 'lognormal'],
    par=[{'loc': 0, 'scale': 1/.9, 'c': 1/.9}, {'loc': 0, 'scale': 10, 'shape': 1.5}],
    ),
    copula=Copula(
    dist='frank',
    par={'par': 1.2, 'dim': 2}
    ),
    n_sim=500000,
    random_state=10,
    n_iter=8
    )
\end{minted}

Besides marginal and copula assumptions, the instantiation of \texttt{LossAggregation} accepts the arguments \texttt{n\_sim} and \texttt{random\_state} to set the number of simulation and the (pseudo)random number generator initialiser of the  Monte Carlo simulation. If \texttt{n\_sim} and the \texttt{random\_state} are omitted the execution is bypassed. Nonetheless, the user can perform it at a later point through the \texttt{dist\_calculate} method. Also, \texttt{n\_iter} controls the number of iterations of the AEP algorithm. This parameter is optional (default value is 7) and can be specified either at the creation of class or directly in methods that include the use of this approach.

In the following code block, we show how to calculate the cdf of the sum of a generalised Pareto and a lognormal dependent random variables, using the AEP algorithm (\texttt{`aep'}) and the Monte Carlo simulation approach (\texttt{`mc'}). The underlying dependency structure is a Frank copula.
\begin{minted}[mathescape]{python}
>>> s = 300 # arbitrary value
>>> p_aep = lossaggregation.cdf(x=s, method='aep')
>>> print('P(X1+X2 <= s) = ', p_aep)
P(X1+X2 <= s) = 0.9811620158197308
>>> p_mc = lossaggregation.cdf(x=s, method='mc')
>>> print('P(X1+X2 <= s) = ', p_mc)
P(X1+X2 <= s) = 0.98126
\end{minted}

\texttt{LossAggregation} includes other functionalities like the survival function \texttt{sf}, the quantile function \texttt{ppf} and the generator of random variates \texttt{rvs}. Furthermore, for the Monte Carlo simulation approach, it is possible to derive empirical statistics and moments using methods such as for example \texttt{moment}, \texttt{mean}, \texttt{var}, \texttt{skewness}, \texttt{censored\_moment}.
To conclude, the last code block illustrates the calculation of the quantile function via the \texttt{ppf} method.

\begin{minted}[mathescape]{python}
>>> lossaggregation.ppf(q=p_aep, method='aep')
300.0000003207744
>>> lossaggregation.ppf(q=p_mc, method='mc')
299.9929982860278
\end{minted}

\subsection{Comparison of the methods for computing the cdf}

In this section, we compared our implementation of the AEP algorithm with the alternative solution based on Monte Carlo simulation in terms of speed and accuracy.
Accuracy in the calculation of the cdf has been assessed by means of the relative error with respect to the reference value for a chosen set of quantiles (i.e. error = estimate/reference - 1).
We replicate the experiment in \citeA{arbenz11} and take the reference values that the authors computed in the original manuscript.
The analysis considers four different Clayton-Pareto models, for $d=2,3,4,5$, with the following parametric assumptions.
In the two-dimensional case ($d=2$), the tail parameters $\gamma$ of the marginal distributions are $0.9$ and $1.8$; the Clayton copula has parameter $\theta=1.2$. In three dimensions ($d=3$), the additional marginal has parameter $\gamma=2.6$, and the copula has $\theta=0.4$.
For the four-dimensional ($d=4$) and five-dimensional ($d=5$) cases, the extra marginal component has parameter equal to $3.3$ and $4$, and the Clayton copula has parameter 0.2 and 0.3, respectively.
The cdf has been evaluated at the quantiles $s=\{10^0, 10^2, 10^4, 10^6\}$ for d = 2 and d = 3, and $s=\{10^1, 10^2, 10^3, 10^4\}$, when d = 4 and d = 5. 

The test of the speed of our implementation has been carried out by measuring the execution time of the cdf function for a single quantile with the built-in \texttt{timeit} library. In line with best practice \cite[Chapter 18]{martelli05}, the observed minimum execution time of independent repetitions of the function call was adopted.

\Cref{table:AEPvsMC} shows the results of our comparison of the two methodologies for the calculation of the cdf.
\begin{table}[!htbp]
\centering
\scalebox{0.9}{
\begin{tabular}{c|r|r|c|r|r}
  \toprule
 \multicolumn{6}{c}{Reference values:} \\
\midrule
  \multicolumn{1}{c}{Quantile} & \multicolumn{1}{c}{d=2} & \multicolumn{1}{c}{d=3} & \multicolumn{1}{c}{Quantile} & \multicolumn{1}{c}{d=4} & \multicolumn{1}{c}{d=5}  \\
\midrule
$s=10^0$ & 0.315835041363441 & 0.190859309689430& $s=10^1$ & 0.983690398913354 & 0.983659549676444
 \\
$s=10^2$ & 0.983690398913354 & 0.983659549676444& $s=10^2$ & 0.983690398913354 & 0.983659549676444\\
$s=10^4$ & 0.999748719229367 & 0.999748708770280& $s=10^3$ & 0.983690398913354 & 0.983659549676444 \\
$s=10^6$ & 0.999996018908404 & 0.999996018515584& $s=10^4$ & 0.983690398913354 & 0.983659549676444 \\
\bottomrule 
\end{tabular}
}

\vspace{2mm}
\begin{tabular}{l|l|r|r|r|r|r|r}
\toprule
 \multicolumn{1}{c}{Dim.} & \multicolumn{1}{c}{Method} & \multicolumn{1}{c}{Time (sec.)}& \multicolumn{1}{c}{Time* (sec.)}&\multicolumn{1}{c}{$s=10^0$}  & \multicolumn{1}{c}{$s=10^2$} & \multicolumn{1}{c}{$s=10^4$} & \multicolumn{1}{c}{$s=10^6$}\\
\midrule
\multirow{8}{1cm}{d=2}
&AEP (7 iter.)& 0.02& 0.01& 4.62e-11& -1.86e-09& 4.13e-08& 1.22e-09\\
&AEP (10 iter.)& 0.06&  0.06&9.03e-14& 5.56e-13& -6.38e-11& 3.88e-11\\
&AEP (13 iter.)& 0.95&  1.61&6.91e-14& 5.04e-13& 1.10e-12& 4.47e-13\\
&AEP (16 iter.)& 24.0&  49.25&-1.72e-13& 4.39e-13& 1.26e-12& 5.66e-13\\
&MC (10**4 sim.)& 0.01&  -&-1.00e-02& -2.13e-04& 4.87e-05& -3.98e-06\\
&MC (10**5 sim.)& 0.03&  -&-8.07e-04& -1.01e-04& 1.87e-05& 1.60e-05\\
&MC (10**6 sim.)& 0.28&  -&-7.90e-05& -3.82e-05& -2.28e-06& 1.02e-06\\
&MC (10**7 sim.)& 2.76& -& 1.71e-04& 6.20e-06& -6.18e-06& -6.81e-07\\
\midrule
\multirow{8}{1cm}{d=3}
&AEP (7 iter.)& 0.04& 0.02& -4.61e-06& -1.15e-06& 1.12e-06& 1.83e-08\\
&AEP (9 iter.)& 0.28&  0.41&-1.73e-07& -3.06e-07& 2.39e-07& 4.26e-09\\
&AEP (11 iter.)& 4.0&  6.65&-6.89e-09& -1.13e-08& 2.95e-08& 7.66e-10\\
&AEP (13 iter.)& 68.4&  118.50&-9.12e-10& -1.29e-09& -6.19e-09& -1.09e-10\\
&MC (10**4 sim.)& 0.01&  -&5.22e-02& -1.43e-04& -5.13e-05& -3.98e-06\\
&MC (10**5 sim.)& 0.04& - &7.59e-03& -3.56e-04& 1.87e-05& -3.98e-06\\
&MC (10**6 sim.)& 0.36& - &-3.87e-03& 1.07e-05& -2.29e-06& -1.98e-06\\
&MC (10**7 sim.)& 4.1& - &7.77e-04& 4.31e-05& 1.61e-06& -4.81e-07\\
\midrule
 \multicolumn{1}{c}{Dim.} &\multicolumn{1}{c}{Method} & \multicolumn{1}{c}{Time (sec.)}& \multicolumn{1}{c}{Time* (sec.)} & \multicolumn{1}{c}{$s=10^1$}  & \multicolumn{1}{c}{$s=10^2$} & \multicolumn{1}{c}{$s=10^3$} & \multicolumn{1}{c}{$s=10^4$}\\
\midrule
\multirow{8}{1cm}{d=4}
&AEP (4 iter.)& 0.05&  0.03&-1.13e-04& 5.03e-04& 7.39e-05& 9.30e-06\\
&AEP (5 iter.)& 0.38&  0.47&-4.45e-04& 1.56e-04& 2.71e-05& 3.42e-06\\
&AEP (6 iter.)& 5.23&  7.15&-4.80e-04& -5.09e-05& -3.69e-06& -4.52e-07\\
&AEP (7 iter.)& 83.55&  107.70&-4.55e-04& -1.56e-04& -2.26e-05& -2.85e-06\\
&MC (10**4 sim.)& 0.01& -&3.18e-03& -1.92e-03& 3.51e-04& 4.42e-04\\
&MC (10**5 sim.)& 0.05& -&6.57e-04& -1.71e-04& 1.03e-05& -7.74e-06\\
&MC (10**6 sim.)& 0.46& -&-6.00e-04& 1.34e-05& -4.75e-06& -2.67e-05\\
&MC (10**7 sim.)& 4.55& -&-2.23e-04& -1.24e-04& -3.29e-05& -5.04e-06\\
\midrule
\multirow{8}{1cm}{d=5}
&AEP (3 iter.)& 0.03& 0.01&-4.72e-03& -5.16e-05& 5.24e-06& 7.22e-07\\
&AEP (4 iter.)& 0.15& 0.20&-6.87e-04& 3.63e-04& 5.30e-05& 6.68e-06\\
&AEP (5 iter.)& 2.66& 4.37&-1.77e-04& 1.94e-04& 2.84e-05& 3.57e-06\\
&AEP (6 iter.)& 65.61& 92.91&-5.14e-13& -9.37e-12& 1.44e-11& -1.17e-11\\
&MC (10**4 sim.)& 0.01& -&5.38e-03& -1.67e-03& -3.70e-04& 1.40e-04\\
&MC (10**5 sim.)& 0.06& -&-7.25e-04& 2.58e-04& -9.29e-06& -9.02e-05\\
&MC (10**6 sim.)& 0.58& -&6.14e-04& -1.89e-04& -1.93e-05& 5.81e-06\\
&MC (10**7 sim.)& 5.65& -&5.29e-04& -2.68e-04& -7.70e-05& -1.02e-05\\
\bottomrule
\end{tabular}
\caption{\label{table:AEPvsMC} Accuracy and speed of the cdf calculation using the AEP algorithm and the Monte Carlo simulation approach (MC) for the sum of Pareto random variables coupled with a Clayton copula for different dimensions and quantiles. The upper table contains the reference values from \protect\citeA{arbenz11}. The lower table reports the execution times in seconds and the relative errors with respect to the reference values. The time column on the left is about the execution times of \textbf{gemact}, while the time column on the right (labelled with a *) lists those of the original manuscript.}
\end{table}

It can be noted that our implementation of the AEP algorithm shows a high accuracy in the calculation of the cdf, for all quantiles and dimensions. Its precision also remains valid for the five-dimensional case. The figures are in line with the results of the original study.
In general, in cases considered, the AEP algorithm is closer to the reference values than the Monte Carlo simulation approach. Nevertheless, the latter shows contained errors whose order of magnitude is $10^{-2}$ at most, when the number of simulation is set to the lowest value.
On the other hand, the AEP algorithm is outperformed by the Monte Carlo simulation approach, in terms of execution speed. The largest gaps have been observed especially when the number of iterations of the AEP is higher and the dimension is $4$ and $5$.
However, it should be noted that the computational times for the AEP algorithm remains acceptable, in most cases below one second even in high dimensions. For the sake of completeness and as a reference, the computational time figures of the first implementation, reported in the study of the original manuscript, are also given.

\begin{figure}[ht]
\centering
\centering\includegraphics[scale=0.52]{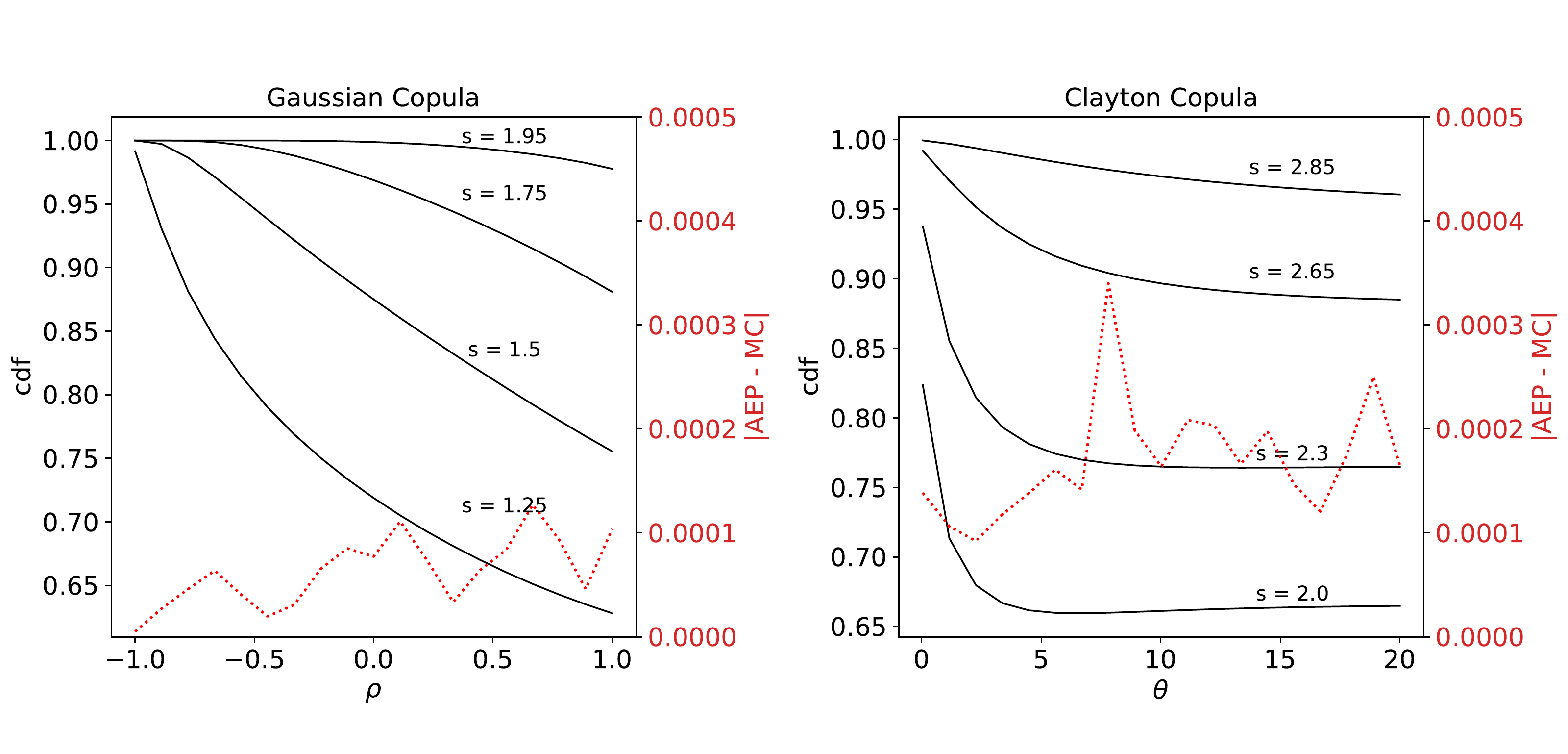}
    \caption{\label{fig:lossaggr_copula_sensitivity} 
    Sensitivity analysis of the cdf at four quantiles $s$ calculated using the AEP algorithm (\texttt{n\_iter} = 7) for different copula models, dimensionality, and underlying degree of dependency. 
    The values of $s$ are 1.25, 1.5, 1.75, 1.95 for the bivariate Gaussian copula (left plot) and 2, 2.3, 2.65, 2.85 for the three-dimensional Clayton (right plot).
    Each solid black line indicates the values of the cdf for a given $s$, as the respective parameters $\rho$ and $\theta$ change.
    The results of the AEP algorithm correspond with those of the Monte Carlo simulation approach (MC), using $10^7$ number of simulations. The dashed red line represents the average absolute difference between the two method cdf values, calculated across the four quantiles.
    }
\end{figure}

To conclude, we perform a sensitivity study of the two above-mentioned approaches for different dependency structures and number of dimensions.
\Cref{fig:lossaggr_copula_sensitivity} shows how the cdf value calculated with the AEP algorithm  changes as the underlying degree of dependency varies, in the cases of a bivariate Gaussian copula and a three-dimensional Clayton copula.
The solid black lines of the graphs represent the cdf evaluated at four quantiles $s$ for increasing values of $\rho$ (the non-diagonal entry of the correlation matrix) and $\theta$ parameters, for the Gaussian copula and the Clayton copula respectively.
The cdf values were also compared with those obtained by the Monte Carlo simulation approach. 
The average absolute difference between the results of the two methods, across the four quantiles, is highlighted by the dotted red line. It can be seen that this remains stable at low values, regardless of the underlying dependency structure. In all cases, the results produced by the two methods almost coincide.  

\section{Loss reserve}
\label{sec:lossreserve}

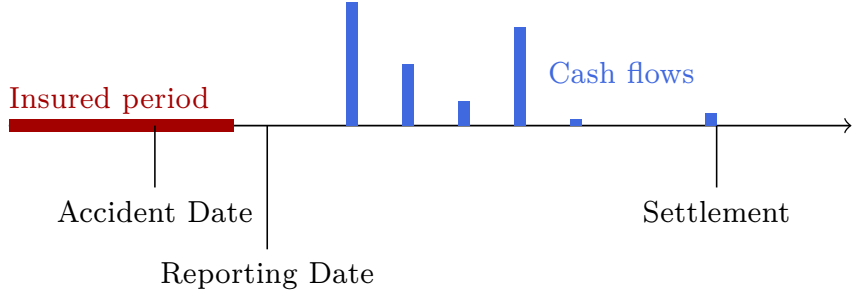
\begin{figure}
    \centering
\resizebox{12cm}{4cm}{%
\begin{tikzpicture}[%
    every node/.style={
        font=\scriptsize,
        text height=1ex,
        text depth=.25ex,
    },
]
\draw[->] (0,0) -- (7.5,0);

\fill[royalblue] (3,0) rectangle (3.1,1);
\fill[royalblue] (3.5,0) rectangle (3.6,.5);
\fill[royalblue] (4,0) rectangle (4.1,.2);
\fill[royalblue] (4.5,0) rectangle (4.6,.8);
\fill[royalblue] (5,0) rectangle (5.1,.05);
\fill[royalblue] (6.2,0) rectangle (6.3,.1);

\node[text width=3cm] at (1.5,.2) 
    {\textcolor{darkcandyapplered}{Insured period}};
    
\node[text width=3cm] at (6.3,.4) 
    {\textcolor{royalblue}{Cash flows}};

\fill[darkcandyapplered] (0,.05) rectangle (2,-0.05);

\draw (1.3,0) -- (1.3,-5mm) node[below] {Accident Date};

\draw (2.3,0) -- (2.3,-10mm) node[below] {Reporting Date};

\draw (6.3,0) -- (6.3,-5mm) node[below] {Settlement};





\end{tikzpicture}%

}    
\caption{Example of events of a non-life insurance claim.}
\label{fig:timeline}
\end{figure}

In non-life insurance, contracts do not settle when insured events occur. At the accident date, the insured event triggers a claim that will generate payments in the future. The task of predicting these liabilities is called \textit{claims reserving} and it assesses claim outstanding loss liabilities \cite<see for example>[p.~11]{wuthrich15}. In the present work, we refer to the total outstanding loss liabilities of past claims as the \textit{loss reserve} or \textit{claims reserve}. \Cref{fig:timeline} sketches an example of the timeline evolution of an individual claim. The insured event occurs within the insured period but the claim settles after several years.
In particular, after the claim is reported, the insurance company makes an initial quantification of the claim payment size, the so-called case estimate.
Two payments occur thereafter. These payments are not known at the evaluation date and they require to be estimated. In between the payments, the case estimate is updated. Until a claim settle, the insurance company refers to it as an open claim. In certain circumstances, settled claims can be reopened \cite[p.~431]{friedland10}.

In this section, we first define the development triangles, the data structure commonly used by actuarial departments for claims reserving \cite[p.~51]{friedland10}.
These are aggregate representations of the individual claim data. 
Then, we present the collective risk model for claims reserving in \citeA{ricotta16} and \citeA{clemente19}, which allows to estimate the variability of the reserve. This model requires extra-inputs from a deterministic model to be implemented. In this manuscript, we rely on the \textit{Fisher-Lange} model \cite{fisher99} as proposed in \citeA{savelliDATA}. Further details can be found in \Cref{app:FL}.

\subsection{Problem framework}

Let the index $i=0,\ldots,\mathcal{J}$ denotes the claim accident period, and let the index $j=0,\ldots,\mathcal{J}$ represents the claim development period, over the time horizon $\mathcal{J}>0$.
The so-called \textit{development triangle} is the set

$$
\mathcal{T}=\{(i, j): \; i=0, \ldots, \mathcal{J}, \; j=0, \ldots, \mathcal{J}; \; i+j \leq \mathcal{J}\}, \quad \mathcal{J}>0.
$$

Below is the list of the development triangles used in this section.

\begin{itemize}
    \item The triangle of \textit{incremental paid claims}:
    $$X^{\mathcal{(T)}} =\left\{x_{i, j}: (i, j) \in \mathcal{T} \right\}, $$
    with $x_{i, j}$ being the total payments from the insurance company for claims occurred at accident period $i$ and paid in period $i+j$.
    \item The triangle of the \textit{number of paid claims}:
    $$N^{\mathcal{(T)}} =\left\{n_{i, j}: (i, j) \in \mathcal{T} \right\}, $$
    with $n_{i, j}$ being the number of claim payments occurred in accident period $i$ and paid in period $i+j$. The triangle of average claim cost can be derived from the incremental paid claims triangle and the number of paid claims. Indeed, we define:
    
    $$m_{i,j}=\frac{x_{i,j}}{n_{i, j}},$$ 
    
    where $m_{i,j}$ is the average claim cost for accident period $i$ and development period $j$.
    \item The triangle of \textit{incremental amounts at reserve}:
    $$R^{\mathcal{(T)}} =\left\{r_{i, j}: (i, j) \in \mathcal{T} \right\}, $$
    with $r_{i, j}$ being the amount booked in period $i+j$ for claims occurred in accident period $i$.
    \item The triangle of the \textit{number of open claims}:
    $$O^{\mathcal{(T)}} =\left\{o_{i, j}: (i, j) \in \mathcal{T} \right\},$$
    with $o_{i, j}$ being the number of claims that are open in period $i+j$ for claims occurred in accident period $i$.
    \item The triangle of the \textit{number of reported claims}:
    $$D^{\mathcal{(T)}} =\left\{d_{i, j}: (i, j)  \in \mathcal{T} \right\}, $$
    with $d_{i, j}$ being the claims reported in period $j$ and belonging to accident period $i$. Often, the number of reported claims is aggregated by accident period $i$. 
\end{itemize}

We implemented the model of \citeA{ricotta16, clemente19}, hereafter referred to as CRMR, which connects claims reserving with aggregate distributions. 
Indeed, the claims reserve is the sum of the (future) payments 

\begin{equation}
\label{eq:claimsreserve}
    R = \sum_{i+j>\mathcal{J}}X_{i,j}.
\end{equation}

The authors represent the incremental payments in each cell of the triangle of incremental payments as a compound mixed Poisson-gamma distribution under the assumptions in \cite{ricotta16}. For $i,j=0,\ldots,\mathcal{J},$
    
\begin{equation}
    \label{eq:crmreserving}
        X_{i,j}=\sum_{h=1}^{N_{i,j}} \psi Z_{h;i,j},
\end{equation}

where the index $h$ is referred to the individual claim severity. The random variable $\psi$ follows a gamma distribution, see \Cref{tab:tabofparslr}, and introduces dependence between claim-sizes of different cells. 

\subsubsection{Predicting the claims reserve}

In order to estimate the value for the claims reserve $R$, this approach requires additional parametric assumptions on $N_{i,j}$ and $Z_{h;i,j}$ when $i+j > \mathcal{T}$. In this section, we illustrate how to use the results from the Fisher-Lange 
to determine the parameters of $N_{i,j}$ and $Z_{h;i,j}$. The Fisher-Lange is a deterministic average cost method for claims reserving \cite[Section H]{ifoa1}. The interested reader can refer to \Cref{app:FL} for a discussion on the Fisher-Lange and an implementation of the model using \textbf{gemact}.

For any cell ($i,j$), with $i+j > \mathcal{T}$, the Fisher-Lange can be used to determine the expected number of future payments $\hat n_{i,j}$ and the future average claim cost $\hat m_{i,j}$. The original concept behind the CRMR is found in \citeA{savelli09}, where the authors assumed that $N_{i,j}$ is Poisson distributed with mean $\hat{n}_{i,j}$ and $Z_{h;i,j}$ is gamma distributed with mean $\hat m_{i,j}$. The coefficient of variation of $Z_{h;i,j}$ is $\hat c_{z_j}$, the relative variation of the individual payments in development period $j$, independent from the accident period. The values for $\hat c_{z_j}$ are estimated using the individual claims data available to the insurer at the evaluation date. 
The CRMR is extended in \citeA{ricotta16} to take into account of the variability of the severity parameter estimation (\textit{Estimation Variance}), in addition to the random fluctuations of the underlying process (\textit{Process Variance}) \cite[p.~28]{wuthrich15}. This is achieved by considering two structure variables ($q$ and $\psi$), on claim count and average cost, to describe parameter uncertainty on $N_{i,j}$ and $Z_{h;i,j}$. The parametric assumptions of the model are summarised in \Cref{tab:tabofparslr}. For $i,j=0,\ldots,\mathcal{J}$, $\hat m_{i,j}$ and $\hat n_{ij}$ are obtained from the computation of the Fisher-Lange.

\begin{table}[ht]
    \centering
    \begin{tabular}{l|l|l|l|l}
         \toprule
          \multicolumn{1}{c}{}
        & \multicolumn{1}{c}{Distribution} &  \multicolumn{1}{c}{Quantity} & \multicolumn{1}{c}{\texttt{a}} & \multicolumn{1}{c}{\texttt{\texttt{scale}}} \\
          \midrule
          $Z_{h;i,j}$& \multirow{3}{2cm}{Gamma}& Individual Payment Cost &$\hat c_{Z_{i,j}}^{-2}$ & $\hat \sigma_{Z_{i,j}}^{2} \hat m_{i,j}$   \\
          $\psi$&& Structure Variable (Individual Payment Cost)&$\hat \sigma_{\psi}^{-2}$ & $\hat \sigma_{\psi}^{2}$  \\
          $q$&& Structure Variable (Payment Number)&$\hat \sigma_{q}^{-2}$ & $\hat \sigma_{q}^{2}$  \\
          
        \bottomrule
    \end{tabular}
    \vspace{1cm}
    
    \begin{tabular}{l|l|l|l|l}
         \toprule
         \multicolumn{1}{c}{}
        & \multicolumn{1}{c}{Distribution} &  \multicolumn{1}{c}{Quantity} & \multicolumn{1}{c}{\texttt{mu}}  \\
          \midrule
          $N_{i,j}$& Mixed Poisson & Payment number & \multicolumn{1}{c}{$\hat n_{i,j} q$ }   \\
        \bottomrule
    \end{tabular}
    \caption{Parametric assumptions of the CRMR. In the upper table we show the parameters for $Z_{h;i,j}, \psi, $ and $q$ that are gamma distributed. The parameters of the structure variables are specified from the user starting from the variance of $\psi$ and $q$, indeed \protect \citeA{ricotta16} assume $\mathbb{E}\left[\psi\right]=\mathbb{E}\left[q\right]=1$. The estimator for the average cost of the individual payments is derived with the Fisher-Lange. The variability of the individual payments is instead obtained from the company database. In the lower table we show the parameter for the claim payment number, i.e. a mixed Poisson-gamma distribution with $\hat n_{i,j}$ derived from the Fisher-Lange.}
    \label{tab:tabofparslr}
\end{table}

\subsection{Illustration \texttt{lossreserve}}

In this section, we show an example of the CRMR using the \texttt{lossreserve} module.
In this respect, we simulated the claims reserving data sets using the individual claim simulator in \citeA{avanzi21} to generate the upper triangles $X^{\left(\mathcal{T}\right)}, N^{\left(\mathcal{T}\right)}, O^{\left(\mathcal{T}\right)}$ and $D^{\left(\mathcal{T}\right)}$. The $R^{\left(\mathcal{T}\right)}$ upper triangle has been simulated using the simulator of \citeA{avanzi23}. The data are simulated using the same assumptions (on an yearly basis) of the \textit{Simple, short tail claims} environment shown in \citeA{almudafer22}. 

No inflation is assumed to simplify the forecast of the future average costs. Estimating and extrapolating a calendar period effect in claims reserving is a delicate and complex subject, a more detailed discussions can be found in \citeA{kuang08b, kuang08, pittarello23}. Furthermore, in practice, insurers might require a specific knowledge of the environment in which the agents operate to set a value for the inflation \cite{avanzi23}. Here, we want to limit the assumptions for this synthetic scenario.

Since the entire claim history is available from the simulation and we know the (true) value for the future payments, we can use this information to back-test the performance of our model \cite{gabrielli19}. In particular, we refer to the actual amount that the insurer will pay in future calendar years as the \textit{actual reserve}, i.e. the amount that the insurer should set aside to cover exactly the future obligations. The CRMR is compared with the \textit{chain-ladder} method of \citeA{mack93}, hereafter indicated as CHL, from the \textbf{R} package \texttt{ChainLadder} \cite{chainladderR}. These data, together with the data sets from \citeA{savelliDATA}, have been saved in the \texttt{gemdata} module for reproducibility. 

\begin{minted}[mathescape]{python}
>>> from gemact import gemdata
>>> ip = gemdata.incremental_payments_sim
>>> pnb = gemdata.payments_number_sim
>>> cp = gemdata.cased_payments_sim
>>> opn = gemdata.open_number_sim
>>> reported = gemdata.reported_claims_sim
>>> czj = gemdata.czj_sim
\end{minted}

\begin{figure}
    \centering
    \begin{tikzpicture}
    \draw (0,0) -- (3,0) -- (3,1.5) -- (0,1.5) -- (0,0); 
    \draw (5+0,0) -- (5+3,0) -- (5+3,1.5) -- (5+0,1.5) -- (5+0,0);
    \draw (2.5+0,-2.5+0) -- (2.5+3,-2.5+0) -- (2.5+3,-2.5+1.5) -- (2.5+0,1.5-2.5) -- (2.5+0,0-2.5);
    \node[align=center] at (1.5,.75) 
    {\texttt{ReserveModel}};
    \node[align=center] at (5+1.5,.75) 
    {\texttt{AggregateData}};
    \node[align=center] at (2.5+1.5,.75-2.5) 
    {\texttt{LossReserve}};
    \draw  (1.5,0) -- (1.5,-.5) ;
    \draw  (5+1.5,0) -- (5+1.5,-.5) ;
    \draw  (1.5,-.5)-- (5+1.5,-.5) ;
    \draw[->] (2.5+1.5,-.5) -- (2.5+1.5,-.9) ;
\end{tikzpicture}
    \caption{Class diagram of the \texttt{lossreserve} module. A rectangle represents a class; an arrow connecting two classes
     indicates that the target class employs the origin class as an attribute.
    In this case, a \texttt{LossReserve} object entails \texttt{ReserveModel} and \texttt{AggregateData} class instances.}
    \label{fig:lossreservediagram}
\end{figure}
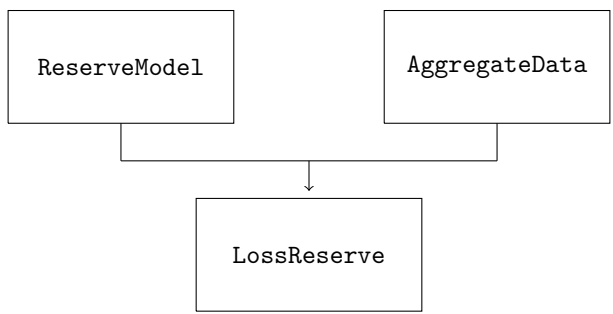

The \texttt{lossreserve} class diagram is illustrated in \Cref{fig:lossreservediagram}: the triangular data sets are stored in the \texttt{AggregateData} class, and the model parameters are contained in the \texttt{ReservingModel} class. The actual computation of the loss reserve is performed with the \texttt{LossReserve} class that takes as inputs an \texttt{AggregateData} object, a \texttt{ReservingModel} object and the parameters of the claims reserving model computation.

\begin{minted}[mathescape]{python}
>>> from gemact import AggregateData
>>> from gemact import ReservingModel
>>> import numpy as np
>>> ad = AggregateData(
    incremental_payments=ip,
    cased_payments=cp,
    open_claims_number=opn,
    reported_claims=reported,
    payments_number=pnb
    )
>>> resmodel_crm = ReservingModel(
    tail=False,
    reserving_method='crm',
    claims_inflation=np.array([1]),
    mixing_fq_par=.01,
    mixing_sev_par=.01,
    czj=czj
    )
\end{minted}

In more detail, the \texttt{ReservingModel} class arguments are:

\begin{itemize}
    \item The \texttt{tail} parameter (boolean) specifying whether the triangle tail is to be modeled.
    \item The \texttt{reserving\_method} parameter, \texttt{`crm'} in this case. 
    \item The \texttt{claims\_inflation} parameter indicating the vector of claims inflation. In this case, as no claims inflation is present, we simply set it to one in all periods.
    \item The mixing frequency and severity parameters \texttt{mixing\_fq\_par} and \texttt{mixing\_sev\_par}. In \citeA{ricotta16} the authors discuss the calibration of the structure variable. Without having a context of a real world example we simply set the structure variables to gamma random variables with mean 1 and standard deviation $0.01$, as it is a medium-low risk value in the authors' examples. 
    \item The coefficients of variation of the individual claim severity computed for each development period. In particular, for each development period, the insurer can compute the coefficient of variation from the individual observations and save them in the vector \texttt{czj}.
    \item The vector \texttt{czj} of the coefficients of variation of the individual claim severity, for each development period. The insurer can compute the coefficient of variation from the individual observations, for each development period. 
\end{itemize}

The computation of the loss reserve occurs within the \texttt{LossReserve} class:

\begin{minted}[mathescape]{python}
>>> from gemact import LossReserve
>>> lr = LossReserve(
    data=ad,
    reservingmodel=resmodel_crm,
    ntr_sim=1000,
    random_state=42
    )
\end{minted}

To instantiate the \texttt{LossReserve} class, \texttt{AggregateData} and \texttt{ReservingModel} objects need to be passed as arguments. The additional arguments required for the computation are: 

\begin{itemize}
    \item the number of simulated triangles \texttt{ntr\_sim},
    \item the (pseudo)random number generator initialiser \texttt{random\_state}.
\end{itemize}

The mean reserve estimate for the CRMR can be extracted from the \texttt{reserve} attribute. In a similarly way to \texttt{LossModel} and \texttt{LossAggregation}, the distribution of the reserve can be accessed from the \texttt{dist} property. The reserve quantiles can be obtained with the \texttt{ppf} method. Output figures are expressed in millions to simplify reading.

\begin{minted}[mathescape]{python}
>>> lr.reserve
8245996481.515498
>>> lr.ppf(q=np.array([.25, .5, .75, .995, .9995]))/10**6
array([8156.79994049, 8244.66099324, 8335.94438221, 8600.42263791, 8676.84206101])
\end{minted}

\subsubsection{Comparison with the chain-ladder}

This section compares the CRMR and the CHL with the actual reserve we know from the simulation. For clarity, the reserve of accident period $i$ is defined as $R_i=\sum^{\mathcal{T}}_{j=\mathcal{T}-i+1} X_{i, j}$. \Cref{tab:ucCL} reports the main results. We can see that the CHL reserve estimate is closer than the CRMR reserve estimate to the actual reserve (the "Reserve" column of each block).
In addition to the reserve estimate, we provide the \textit{mean squared error of prediction} (MSEP) \cite[p.~268]{wuthrich15} of the reserve estimator for each accident period and for the total reserve. In the aforementioned manuscript, the author introduces the MSEP as a measure of risk under the notion in \citeA[p.~169]{wuthrich15} conditional on the information available in the upper triangle. The MSEP can be decomposed into two components, the irreducible risk arising from the data generation process (process error, PE) and a risk component arising from the model used to calculate the reserve (model error, ME). We can provide a reference value for the true PE by running multiple simulations from the simulator described in the previous section and calculating the standard error of the reserves by accident period. To obtain the PE, we simulated $100$ triangles. Net of the ME component, the results in \Cref{tab:ucCL} appear to show that the CHL underestimates the forecast error in the early accident periods and overestimates the forecast error in the later accident periods. 

\begin{table}[H]
\centering
\begin{adjustbox}{max width=\textwidth}
\begin{tabular}{l|rr|rr|rr}
\toprule
 \multicolumn{1}{c}{Accident} & \multicolumn{2}{c}{CRMR} & \multicolumn{2}{c}{CHL} & \multicolumn{2}{c}{Actual} \\
 \multicolumn{1}{c}{Period} & \multicolumn{1}{c}{Reserve} & \multicolumn{1}{c}{MSEP} & \multicolumn{1}{c}{Reserve} & \multicolumn{1}{c}{MSEP} & \multicolumn{1}{c}{Reserve} & \multicolumn{1}{c}{PE} \\
\midrule
    0 & 0.00 & 0.00 & 0.00 & 0.00 & 0.00&0.00  \\ 
      1 & 404.30 & 14.37 & 161.24 & 0.003  & 172.03 &7.78 \\ 
      2 & 488.27 &  15.11 & 337.34 & 0.17 & 327.99&11.29   \\ 
      3 & 645.25 & 18.62 & 532.44 & 11.58 & 539.04&17.94 \\ 
      4 &  795.79& 20.34 & 785.44 & 27.40 & 754.93 &20.56 \\ 
      5 &1026.94 & 25.16 &  1134.47 & 40.79 & 1090.84 &20.73 \\ 
      6 & 1303.70&  29.09 & 1398.11 &52.23  & 1464.93 &28.87 \\ 
      7 & 1618.36 & 33.73& 1860.56 & 70.53& 1867.04 &31.95 \\ 
      8 & 1963.40 & 39.51 &  2214.72 & 176.97 & 2382.24&32.70  \\
Total & 8246.00 & 130.09 & 8424.31 & 169.79& 8599.04 & 64.32\\
\bottomrule
\end{tabular}
\end{adjustbox}
\caption{\label{tab:ucCL} Reserve and mean squared error of prediction (MSEP) by accident period for the CRMR and the CHL. The actual reserve and its process error (PE) by accident period are also indicated. Amounts are shown in millions.}
\end{table} 

To conclude, \Cref{tab:finalCL} shows the total reserves of the CRMR and the CHL. In particular, it highlights the reserve amount as a percentage of the actual reserve (8599.04) and the coefficient of variation (CoV), i.e. the MSEP divided by the reserve. The total relative variability referred to the PE component gives a CoV of 0.76 \%.

\begin{table}[H]
\centering
\begin{tabular}{l|l|l|l}
\toprule 
\multicolumn{1}{c}{}& \multicolumn{1}{c}{Reserve} & \multicolumn{1}{c}{Reserve/Actual} & \multicolumn{1}{c}{CoV} \\
\midrule
CRMR & 8246.00 & 0.96 & 1.58\%\\
CHL & 8424.31 & 0.98 & 2.02 \%\\
\bottomrule
\end{tabular}

\caption{\label{tab:finalCL} Total reserve estimates, their relative value, as a fraction of the actual value (8599.04), and their coefficients of variation (CoV), for the CRMR and the CHL. Absolute amounts are reported in millions. }
\end{table}

\section{Conclusions}

This paper introduces \textbf{gemact}, a \textbf{Python} package for non-life actuarial modelling based on the collective risk theory.

Our library expands the reach of actuarial sciences within the growing community of \textbf{Python} programming language. It provides new functionalities and tools for loss modelling, loss aggregation and loss reserving, such as the $(a, b, 0)$ and $(a, b, 1)$ classes of distributions, the AEP algorithm and the collective risk model for claims reserving. Hence, it can be applied in different areas of actuarial sciences, including pricing, reserving and risk management.
The package has been designed primarily for the academic environment. Nevertheless, its use as support for insurance business specialists in prototypes modelling, business studies and analyses is not to be excluded.

The structure of our package aims to ensure modifiability, maintainability and scalability, as we thought of \textbf{gemact} as an evolving and growing project in terms of introducing features, integrating functionalities, enhancing methodologies, and expanding its scopes.
Possible future enhancements could involve the introduction of new probability distribution families, the implementation of supplementary methodologies for the approximation of quantiles of the sum of random variables, and the addition of costing procedures for exotic and non-traditional reinsurance solutions.

\subsection*{Device specifications} 

All experiments and analyses were run on a computer with an Intel\textsuperscript{\textregistered} Core\texttrademark\ i7-1065G7 CPU processor with 16GB RAM, running at 1.30GHz, on Windows 11 Home edition.

\section*{Acknowledgments}

Previous versions of \textbf{gemact} were presented at the Mathematical and Statistical Methods for Actuarial Sciences and Finance 2022, and at the Actuarial Colloquia 2022, in the ASTIN section. We would like to thank all the people who gave us feedback and suggestions about the project. 

\section*{Data Availability Statement}

The data and code supporting the findings of this study are openly available in GitHub at \href{https://github.com/gpitt71/gemact-code}{gpitt71/gemact-code}. Supplementary material not included as code blocks in the manuscript can be found in the subfolder \href{https://github.com/gpitt71/gemact-code/tree/main/gemact/vignette}{vignette}.
The results contained in the manuscript are reproducible, excluding environment-specific numerical errors. These discrepancies do not affect the overall validity of the results.
The \href{https://github.com/gpitt71/gemact-code}{gpitt71/gemact-code} folder is registered with the unique Zenodo DOI reference number \href{https://doi.org/10.5281/zenodo.10117505}{10.5281/zenodo.10117505}.

\bibliography{references}

\newpage

\begin{appendix}

\section{List of the supported distributions} \label{app:alldists}

\Cref{tab:alldists} gives an overview of the distributions available in \textbf{gemact}. In particular, the table presents the distribution name (column one) and its name within \textbf{gemact} apparatus (column two). Moreover, it shows the distribution support (column three) and whether the distribution is supported in \textbf{scipy} (column four). 

\begin{table}[H]
\centering
\begin{tabular}{llll}
\hline
\textbf{Distribution}& \textbf{gemact} \textbf{name}  & \textbf{Support}    & \textbf{scipy} \\ \hline
Binomial&\texttt{binom} & discrete &   $\checkmark$   \\   
 \hline
Geometric&\texttt{geom} & discrete &  $\checkmark$   \\   
 \hline
  Log-Series&\texttt{logser} &discrete  & $\checkmark$   \\   
 \hline
Negative Binomial&\texttt{nbinom}  &discrete   &  $\checkmark$   \\   
 \hline
 Poisson&\texttt{poisson} &discrete  &  $\checkmark$   \\   
 \hline
 PWC&\texttt{pwc} & discrete  &   \textcolor{red}{\ding{55}}  \\   
 \hline
  Zero-Modified Poisson&\texttt{zmpoisson} &discrete&  \textcolor{red}{\ding{55}}  \\
 \hline
 Zero-Modified Binomial&\texttt{zmbinom} & discrete  &  \textcolor{red}{\ding{55}}   \\   
 \hline
Zero-Modified Geometric&\texttt{zmgeom} & discrete &  \textcolor{red}{\ding{55}}   \\   
 \hline
  Zero-Modified Log-Series&\texttt{zmlogser} &discrete & \textcolor{red}{\ding{55}}   \\  
 \hline
Zero-Modified Negative Binomial&\texttt{zmnbinom} & discrete &   \textcolor{red}{\ding{55}}   \\            
 \hline
 Zero-Truncated Binomial&\texttt{ztbinom}  &discrete &  \textcolor{red}{\ding{55}}  \\   
 \hline
Zero-Truncated Geometric& \texttt{ztgeom} & discrete  &   \textcolor{red}{\ding{55}}   \\   
 \hline
Zero-Truncated Negative Binomial&\texttt{ztnbinom} &discrete &   \textcolor{red}{\ding{55}}   \\  
 \hline
 Zero-Truncated Poisson &\texttt{ztpoisson} & discrete &  \textcolor{red}{\ding{55}}   \\   
 \hline
 Beta&\texttt{beta} &continuous  &   $\checkmark$   \\   
 \hline
  Burr & \texttt{burr12} &continuous  &  $\checkmark$   \\        
 \hline
Exponential&\texttt{exponential} & continuous &  $\checkmark$   \\       
 \hline
 Fisk&\texttt{fisk} & continuous  &  $\checkmark$   \\   
 \hline
  Gamma&\texttt{gamma} & continuous &  $\checkmark$   \\       
 \hline
  Generalised Beta&\texttt{genbeta} &continuous  &  \textcolor{red}{\ding{55}}  \\
 \hline
  Generalised Pareto& \texttt{genpareto} &continuous  &  $\checkmark$   \\   
 \hline
  Inverse Burr (Dagum)&\texttt{dagum}&  continuous &  $\checkmark$   \\   
 \hline
 Inverse Gamma&\texttt{invgamma} & continuous &  $\checkmark$   \\
  \hline
  Inverse Gaussian&\texttt{invgauss} & continuous  &  $\checkmark$   \\   
 \hline
  Inverse Paralogistic &\texttt{invparalogistic} & continuous &   \textcolor{red}{\ding{55}}   \\  
 \hline
Inverse Weibull&\texttt{invweibull} &continuous  &  $\checkmark$   \\    
\hline 
Log Gamma&\texttt{loggamma} & continuous &  $\checkmark$   \\
 \hline
 Lognormal& \texttt{lognormal} &continuous  &  $\checkmark$   \\  
 \hline
  Paralogistic & \texttt{paralogistic}&continuous &  \textcolor{red}{\ding{55}}  \\ 
\hline
 Pareto (One-Parameter)& \texttt{pareto1} &continuous  &  \textcolor{red}{\ding{55}}   \\   
 \hline
 Pareto (Two-Parameter)& \texttt{pareto2} &continuous  &  $\checkmark$   \\   
 \hline
 PWL&\texttt{pwl} & continuous  &   \textcolor{red}{\ding{55}}   \\   
 \hline
 Uniform&\texttt{uniform} & continuous  &  $\checkmark$   \\  
 \hline
Weibull&\texttt{weibull}  & continuous &  $\checkmark$   \\   
 \hline
\end{tabular}
\caption{\label{tab:alldists} List of distributions supported by \textbf{gemact}.}
\end{table}

Lastly, the list of available copulas is provided in \Cref{tab:allcopulas}.

\begin{table}[H]
\centering
\begin{tabular}{lll}
\hline
\textbf{Copula} & \textbf{gemact} \textbf{name}  & \textbf{Family} \\
\hline
Ali-Mikhail-Haq&\texttt{ali-mikhail-haq} &  Archimedean\\
\hline
Clayton&\texttt{clayton} & Archimedean \\  
\hline
Frank&\texttt{frank} & Archimedean \\ 
\hline
Gumbel&\texttt{gumbel} &  Archimedean\\
\hline
Joe&\texttt{joe} &  Archimedean\\
\hline
Gaussian&\texttt{gaussian} & Elliptical \\ 
\hline
Student t&\texttt{tstudent} & Elliptical \\ 
\hline
Fréchet-Hoeffding lower bound (W)&\texttt{frechet–hoeffding-upper}& Fundamental \\
\hline
Fréchet-Hoeffding upper bound (M)&\texttt{frechet–hoeffding-lower}& Fundamental \\
\hline
Independent&\texttt{independent} & Fundamental \\
\hline
\end{tabular}
\caption{\label{tab:allcopulas} List of copulas supported by \textbf{gemact}.}
\end{table}

\section{The AEP algorithm}
\label{app:AEP}

The AEP algorithm is a numerical procedure, based on a geometric approach, to calculate the joint cdf of the sum of dependent random variables.
The graphical interpretation of the basic idea of this algorithm is straightforward, especially in the two-dimensional case. Therefore, to facilitate the reader's understanding, in this section we limit ourselves to the case where $d = 2$, and focus only on its first two iterations. The underlying logic can then be extended to $d \geq 2$ and to any number of iterations.

Let us first define, for $b_1,b_2 \in \mathbb R$, a simplex as:

\begin{equation*}
\mathcal{S}((b_1, b_2), h)= \begin{cases}\left\{x_1,x_2 \in \mathbb{R}: x_1-b_1>0, x_2-b_2>0, \text { and } \sum_{k=1}^2\left(x_k-b_k\right) \leq h\right\} & \text { if } h>0, \\ \left\{x_1,x_2 \in \mathbb{R}: x_1-b_1 \leq 0, x_2-b_2 \leq 0, \text { and } \sum_{k=1}^2\left(x_k-b_k\right)>h\right\} & \text { if } h<0 ,\end{cases}
\end{equation*}

and a square as:

\begin{equation*}
\mathcal{Q}((b_1,b_2), h)= \begin{cases}\left(b_1, b_1+h\right] \times\left(b_2, b_2+h\right] & \text { if } h>0, \\ \left(b_1+h, b_1\right] \times\left(b_2+h, b_2\right] & \text { if } h<0 .\end{cases}
\end{equation*}

The $H$-measure of the square $V_H\left(\mathcal{Q}(b_1,b_2, h)\right)$ is computed as follows \cite[p.~8]{nelsen07}:

\begin{equation*}
V_H\left(\mathcal{Q}((b_1,b_2), h)\right)  = H\left(b_1+h, b_2+h\right)-H\left(b_1, b_2+h\right)-H\left(b_1+h, b_2\right)+H\left(b_1, b_2\right),
\end{equation*}

where $H$ the joint cdf in \Cref{eq:simplexprob}.
In general, the algorithm is based on the observation that a simplex approximated by a square generates three smaller simplexes, each of which can in turn be approximated by a square that generates three new, even smaller simplexes, and so on.
By repeating this iterative scheme with an increasing number of iterations, the quality of the approximation improves and the error tends to 0.
It can be noted that some simplexes generated by the process lies outside the original simplex. The measure of those needs to be subtracted instead of being added.

\begin{figure}[!ht]
\centering
\begin{subfigure}[t]{0.3\textwidth}
\centering
\resizebox{\linewidth}{!}{
\begin{tikzpicture}
\draw[->, line width=0.01mm] (0,0)--(4,0) node[right]{$x_1$};
\draw[->, line width=0.01mm] (0,0)--(0,4) node[above]{$x_2$};
\draw[ultra thick](0,3)--(3,0);
\draw[ultra thick](0,0)--(3,0);
\draw[ultra thick](0,0)--(0,3);
\node[text width=1cm] at (0,3-.01){$s$};
\node[text width=.5cm] at (3,0-.4) 
    {$s$};

\filldraw[draw=black, fill=gray!20] (0,3) -- (3,0) -- (0,0) -- cycle;

\node[text width=1cm] at (1,1) 
    {$\mathcal{S}_1$};
\end{tikzpicture}}

\subcaption{\label{fig:simplex} We are interested in $P\left[X_1+X_2 \leq s\right]$. $X_1+X_2 \leq s$ with $s \in \mathbb{R}^+$ is the simplex $\mathcal{S}_1$.}
\end{subfigure}
\hfill
\begin{subfigure}[t]{0.3\textwidth}
\centering
\resizebox{\linewidth}{!}{
\begin{tikzpicture}
\filldraw[draw=black, fill=gray!20] (0,0) -- (2,0) -- (2,2) -- (0,2) -- cycle;
\draw[->, line width=0.01mm] (0,0)--(4,0) node[right]{$x_1$};
\draw[->, line width=0.01mm] (0,0)--(0,4) node[above]{$x_2$};
\draw(0,3)--(3,0);
\draw[ultra thick](2,2)--(0,2);
\draw[ultra thick](0,2)--(0,0);
\draw[ultra thick](0,0)--(2,0);
\draw[ultra thick](2,0)--(2,2);
\node[text width=1cm] at (0,3-.01){$s$};
\node[text width=.5cm] at (3,0-.4) 
    {$s$};

\node[text width=1cm] at (1,1) 
    {$\mathcal{Q}_1$};
\node[text width=1cm] at (2.6,.3) 
    {$\mathcal{S}_2$};
\node[text width=1cm] at (1.95,1.65) 
    {$\mathcal{S}_3$};
\node[text width=1cm] at (.6,2.3) 
    {$\mathcal{S}_4$};

\end{tikzpicture}}

\subcaption{\label{fig:simplex2} Iteration $1$, the simplex $\mathcal{S}_1$ in the first iteration is approximated with the square $\mathcal{Q}_1$. }
\end{subfigure}
\hfill
\begin{subfigure}[t]{0.3\textwidth}
\centering
\resizebox{\linewidth}{!}{
\begin{tikzpicture}

\draw[->, line width=0.01mm] (0,0)--(4,0) node[right]{$x_1$};
\draw[->, line width=0.01mm] (0,0)--(0,4) node[above]{$x_2$};

\filldraw[draw=black, fill=gray!20] (2,2/3) -- (2+2/3,2/3) -- (2+2/3,0) -- (2,0) -- cycle;
\filldraw[draw=black, fill=gray!20] (2,2) --(2-2/3,2) -- (2-2/3,2-2/3) --(2,2-2/3)  -- cycle;
\filldraw[draw=black, fill=gray!20] (0,2+2/3) --(0,2) -- (2/3,2) --(2/3,2+2/3)  -- cycle;

\draw(0,3)--(3,0);
\draw(2,2)--(0,2);
\draw(0,2)--(0,0);
\draw(0,0)--(2,0);
\draw(2,0)--(2,2);

\draw[ultra thick](2,2/3)--(2+2/3,2/3);
\draw[ultra thick](2+2/3,2/3)--(2+2/3,0);
\draw[ultra thick](2,2/3)--(2,0);
\draw[ultra thick](2,0)--(2+2/3,0);

\draw[ultra thick](2/3,2)--(2/3,2+2/3);
\draw[ultra thick](2/3,2+2/3)--(0,2+2/3);
\draw[ultra thick](2/3,2)--(0,2);
\draw[ultra thick](0,2)--(0,2+2/3);

\draw[ultra thick](2,2)--(2-2/3,2);
\draw[ultra thick](2,2)--(2,2-2/3);
\draw[ultra thick](2-2/3,2-2/3)--(2,2-2/3);
\draw[ultra thick](2-2/3,2-2/3)--(2-2/3,2);

\node[text width=1cm] at (0,3-.01){$s$};
\node[text width=.5cm] at (3,0-.4) 
    {$s$};
\node[text width=1cm] at (2.6,.3) 
    {$\mathcal{Q}_2$};
\node[text width=1cm] at (1.95,1.65) 
    {$\mathcal{Q}_3$};
\node[text width=1cm] at (.6,2.3) 
    {$\mathcal{Q}_4$};

\end{tikzpicture}}

\subcaption{\label{fig:simplex3} Iteration $2$, the smaller simplexes $\mathcal{S}_2$, $\mathcal{S}_3$, and $\mathcal{S}_4$ are approximated with the squares $\mathcal{Q}_2, \mathcal{Q}_3, \mathcal{Q}_4$ and their area is added (or subtracted) to obtain $\mathcal{S}_1$. }
\end{subfigure}
\caption{Sketch of the first two iterations of the AEP algorithm in the two-dimensional case. }
\end{figure}
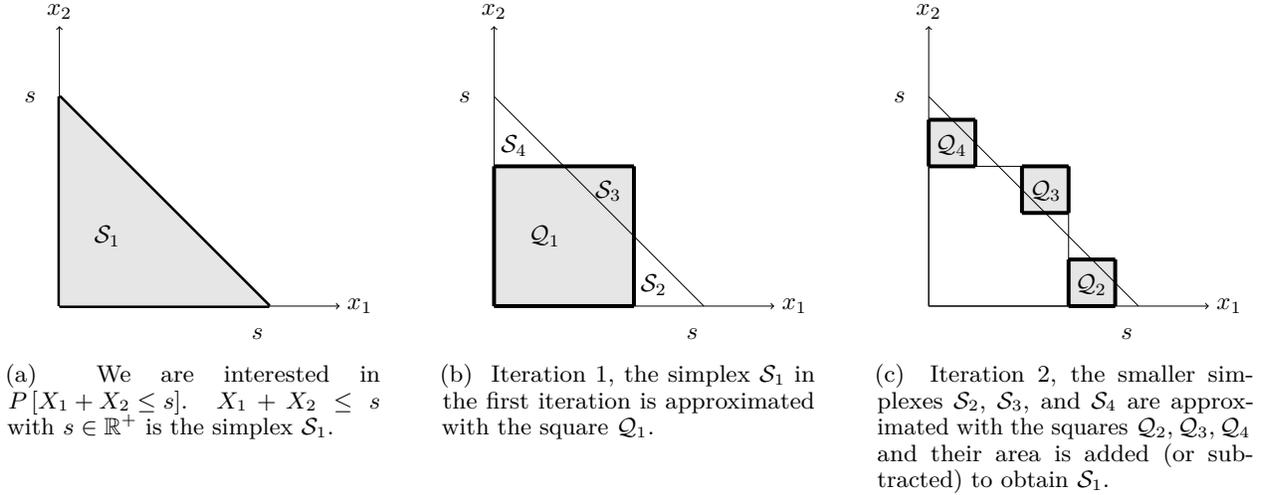

\Cref{fig:simplex} shows the simplex $\mathcal{S}_1=\mathcal{S}((0, 0), s)$, where $s \in \mathbb{R}^+$ is the value at which the joint cdf is calculated in \Cref{eq:simplexprob}.
For the first iteration (\Cref{fig:simplex2}) we adopt the square $\mathcal{Q}_1 = \mathcal{Q}((0,0), \frac 2 3 s)$. \citeA{arbenz11} explains that that the choice of the $2/3$ factor provides fastest convergence when $d=2$. This factor is set automatically in our implementation. Hence, at the end of the first iteration, we have

\begin{equation*}
P\left[X_1+X_2 \leq s \right] \approx P_1 \left(s\right),
\end{equation*}

where $P_1 \left(s\right)=V_H\left(\mathcal{Q}_1\right)$.
For example, if we consider uniform marginals and a Gaussian copula with correlation $0.7$, we would obtain the following.

\begin{minted}[mathescape]{python}
>>> from gemact import Margins, Copula
>>> margins = Margins(
    dist=['uniform', 'uniform'],
    par=[{'a': 0, 'b': 1}, {'a': 0, 'b': 1}]
    )
>>> copula = Copula(
    dist='gaussian',
    par={'corr': [[1, 0.7], [0.7, 1]]}
    )
>>> la = LossAggregation(
    copula=copula,
    margins=margins
    )
>>> la.cdf(x=1, n_iter=1, method='aep')
0.55188934403716
\end{minted}

In the second iteration of the algorithm, shown in in \Cref{fig:simplex3}, we use again the same logic and approximate the simplexes in \Cref{fig:simplex2}:
\begin{align*}
&\mathcal{S}_2 = \mathcal{S} ((\frac 2 3 s, 0) , \frac 1 3 s) \\
&\mathcal{S}_3=\mathcal{S}((\frac 2 3 s, \frac 2 3 s),\frac {-1} {3} s)\\
&\mathcal{S}_4=\mathcal {S}((0,\frac 2 3 s), \frac 1 3 s)\\
\end{align*}
with the squares
\begin{align*}
&\mathcal{Q}_2 = \mathcal{Q} ((\frac 2 3 s, 0) , \frac 1 3 s) \\
&\mathcal{Q}_3=\mathcal{Q}((\frac 2 3 s, \frac 2 3 s),\frac {-1} {3} s)\\
&\mathcal{Q}_4=\mathcal {Q}((0,\frac 2 3 s), \frac 1 3 s).\\
\end{align*}
Note that this time the $H$-measure of $\mathcal{Q}_3$ is subtracted. Similarly to the first iterations the $\frac{1}{3}$ factors are chosen accordingly to the guidelines of the original manuscript to guarantee the fastest convergence.
We obtain, at the second iteration:

\begin{equation*}
P\left[X_1+X_2 \leq s \right] \approx P_2 \left(s\right),
\end{equation*}

with $P_2 \left(s\right)= P_1\left(s\right)+V_H\left(\mathcal{Q}_2\right)-V_H\left(\mathcal{Q}_3\right)+V_H\left(\mathcal{Q}_4\right). $

To conclude, continuing with the previous example, the results for the first two iterations is given in the code block below.
\begin{minted}[mathescape]{python}
>>> la.cdf(x=1, n_iter=2, method='aep')
0.4934418427652146
\end{minted}

\section{Claims reserving with the Fisher-Lange}

\label{app:FL}

This section briefly introduces the Fisher-Lange approach. The claims reserve is the sum of the (future) payments, forecast as the product between the predicted future average cost and the predicted number of future payments, for each cell. In formula:

\begin{equation}
R = \sum_{i+j>\mathcal{J}}\hat m_{i,j} \hat n_{i,j}.
\end{equation}

The average claim cost in the lower triangle is forecast as the projection of the inflated average claim cost. 
\begin{equation}
\widehat{m}_{i, j}=m_{\mathcal{J}-j, j} \prod_{h=\mathcal{J}+1}^{i+j}\left(1+\delta_h\right),
\end{equation}
where $\delta_h$ represents the claims inflation for calendar period $h$.

As far the number of claims are concerned, this method assumes that the future number of paid claims is related to the percentage of open claims at the evaluation date and to the claims settlement speed. Indeed, at the evaluation date, the lower triangle is estimated as:
\begin{equation}
\hat{n}_{i, j}=o_{i, \mathcal{J}-i} \cdot \alpha_{\mathcal{J}-i} \cdot v_j^{(i)},
\end{equation}
where $\left[\alpha_j\right]$, $j=0,\ldots, \mathcal{J}-1$, is the vector of open claims given by $\alpha_j = \mathbb{E}\left[ \tau_{i, j}\right]$, and
\begin{equation}
\tau_{i, j}=\frac{\sum_{h=j+1}^{\mathcal{J}-i} n_{i, h}+o_{i, \mathcal{J}-i}}{o_{i, j}},
\end{equation}
    
for $i=0, \ldots, \mathcal{J}-1$ and $j=0, \ldots, \mathcal{J}-i-1$. It is assumed that $\alpha_\mathcal{J}=1$.
The claim settlement speed is then computed for each accident year. The settlement speed for accident period $\mathcal{J}$ is
    
\begin{equation}
v_j^{(\mathcal{J})}=\frac{n_{\mathcal{J}-j, j} \cdot \frac{d_\mathcal{J}}{d_{\mathcal{J}-j}}}{\sum_{j=1}^\mathcal{J} n_{\mathcal{J}-j, j} \cdot \frac{d_\mathcal{J}}{d_{\mathcal{J}-j}}},
\end{equation}

where $d_i$ represents the number of reported claims for accident period $i$, with $i=0,\ldots,\mathcal{J}$. The formula is corrected for other accident years following the approach in \citeA[p.~141]{savelliDATA}.

Similarly to the CRMR described in \Cref{sec:lossreserve}, the results for the Fisher-Lange can be computed with the \textbf{gemact} package. Below, we show an example using the simulated data sets from \Cref{sec:lossreserve}. 

\begin{minted}[mathescape]{python}
>>> from gemact import gemdata
>>> ip = gemdata.incremental_payments_sim
>>> pnb = gemdata.payments_number_sim
>>> cp = gemdata.cased_payments_sim
>>> opn = gemdata.open_number_sim
>>> reported = gemdata.reported_claims_sim
>>> czj = gemdata.czj_sim
>>> claims_inflation = np.array([1]) 
\end{minted}

The data are represented in the \texttt{AggregateData} class.

\begin{minted}[mathescape]{python}
>>> from gemact.lossreserve import AggregateData
>>> ad = AggregateData(
    incremental_payments=ip,
    cased_payments=cp,
    open_claims_number=opn,
    reported_claims=reported,
    payments_number=pnb)
\end{minted}

Afterwards, we specify the \texttt{ReservingModel}. In this example, we fix the parameter \texttt{tail} to \texttt{True} to obtain an estimate of the tail.

\begin{minted}[mathescape]{python}
>>> resmodel = ReservingModel(
    tail=False,
    reserving_method='fisher_lange',
    claims_inflation=claims_inflation)
\end{minted}

Thereafter, the actual computation of the loss reserve is performed within the \texttt{LossReserve} class:

\begin{minted}[mathescape]{python}
>>> from gemact.lossreserve import LossReserve
>>> lossreserve = LossReserve(data=ad, reservingmodel=resmodel)
\end{minted}

The \texttt{LossReserve} class comes with a summary view of the estimated reserve per each accident period, in a similar way to the \texttt{LossModel} class, the \texttt{print\_loss\_reserve} method. In \Cref{tab:ucCRMRFL} we report the CRMR results from \Cref{tab:ucCL} and we add the results for the Fisher-Lange.

\begin{table}[H]
\centering
\begin{adjustbox}{max width=\textwidth}
\begin{tabular}{l|rr|r|r}
\toprule
 \multicolumn{1}{c}{Accident} & \multicolumn{2}{c}{CRMR} & \multicolumn{1}{c}{Fisher-Lange} & \multicolumn{1}{c}{Actual}\\
 \multicolumn{1}{c}{Period} & \multicolumn{1}{c}{Reserve} & \multicolumn{1}{c}{MSEP} & \multicolumn{1}{c}{Reserve} & \multicolumn{1}{c}{Reserve}  \\
\midrule
    0 & 0.00 & 0.00 & 0.00  & 0.00\\ 
      1 & 404.30 & 14.37 & 404.44  & 172.03 \\ 
      2 & 488.27 &  15.11 & 488.38 & 327.99 \\ 
      3 & 645.25 & 18.62 & 645.98 & 539.04 \\ 
      4 &  795.79& 20.34 &  795.43 & 754.93 \\ 
      5 &1026.94 & 25.16 &  1026.39  & 1090.84 \\ 
      6 & 1303.70&  29.09 & 1302.69 & 1464.93 \\ 
      7 & 1618.36 & 33.73& 1616.50 & 1867.04 \\ 
      8 & 1963.40 & 39.51 &  1962.83 & 2382.24 \\ 
Total & 8246.00 & 130.09 & 8242.64 &  8599.04 \\
\bottomrule
\end{tabular}
\end{adjustbox}
\caption{\label{tab:ucCRMRFL} Reserves by accident period for the CRMR and the Fisher-Lange. We also report the actual reserve. Amounts are shown in millions.}
\end{table} 

As expected, being the Fisher-Lange the underlying methodology to the CRMR the results for the claims reserve provided from the two approaches are consistent. 

On top of this, insights on the behaviour of Fisher-Lange $\left[\alpha_j\right]$ and settlement speed $\left[v^{(i)}_j\right]$, for $j=0,\ldots, \mathcal{J}-1$, can be inspected with the \texttt{plot\_alpha\_fl} and \texttt{plot\_ss\_fl} methods. 

\end{appendix}

\end{document}